\documentclass{article}

\usepackage{algorithm}
\usepackage{algorithmic}

\usepackage[preprint,nonatbib]{arxiv}
\usepackage{wrapfig}
\usepackage{caption}  

\usepackage{amsmath}
\usepackage{amssymb}
\usepackage{mathtools}
\usepackage{amsthm}

\usepackage[utf8]{inputenc} 
\usepackage[T1]{fontenc}    
\usepackage{hyperref}       
\usepackage{url}            
\usepackage{booktabs}       
\usepackage{amsfonts}       
\usepackage{graphicx}
\usepackage{subfigure}
\usepackage{multirow} 

\usepackage{nicefrac}       
\usepackage{microtype}      
\usepackage{xcolor}         
\usepackage{xspace}
\usepackage{float}

\newcommand{\ourmethod}{ViMo\xspace}
\newcommand{\dz}[1]{{\color{black}#1}}

\usepackage{tcolorbox}
\definecolor{beigebg}{RGB}{253,249,238} 
\definecolor{argcolor}{RGB}{70,130,180}
\tcbset{
  mypromptstyle/.style={
    colback=beigebg,
    colframe=beigebg,
    boxrule=0pt,
    sharp corners,
    fontupper=\ttfamily\scriptsize,
    before=\vspace{0.5em},
    after=\vspace{0.5em},
  }
}

\title{A Generative Visual GUI World Model for App Agents}

\author{
  \textbf{Dezhao Luo\textsuperscript{1,*},} 
  \textbf{Bohan Tang\textsuperscript{2,*},} 
  \textbf{Kang Li\textsuperscript{2},}
  \textbf{Georgios Papoudakis\textsuperscript{3},}
  \textbf{Jifei Song\textsuperscript{3},}
  \textbf{Shaogang Gong\textsuperscript{1},} \\
\textbf{Jianye Hao\textsuperscript{3},} \textbf{Jun Wang\textsuperscript{4},} \textbf{Kun Shao\textsuperscript{3,†}}
\\
  \textsuperscript{1}Queen Mary University of London, \textsuperscript{2}University of Oxford, \textsuperscript{3}Huawei Noah's Ark Lab, \\ \textsuperscript{4}University College London \\
}

\begin{document}
\renewcommand{\thefootnote}{\fnsymbol{footnote}}
\footnotetext{* Equal Contribution, † Corresponding author: shaokun2@huawei.com}
\renewcommand{\thefootnote}{\arabic{footnote}}

\maketitle

\begin{abstract}
App agents, which autonomously operate mobile Apps through Graphical User Interfaces (GUIs), have gained significant interest in real-world applications. Yet, they often struggle with long-horizon planning, failing to find the optimal actions for complex tasks with longer steps. To address this, world models are used to predict 
the next GUI observation based on user actions, enabling more effective agent planning. However, existing world models primarily focus on generating only textual descriptions, lacking essential visual details.
To fill this gap, we propose \textbf{\ourmethod}, the first \dz{\textbf{Vi}}sual world  \dz{\textbf{Mo}}del designed to generate future App observations as images.
For the challenge of generating text in image patches, where even minor pixel errors can distort readability, we decompose GUI generation into graphic and text content generation. We propose a novel data representation, the Symbolic Text  Representation~(\textbf{STR}), to overlay text content with symbolic placeholders while preserving graphics. With this design, 
\ourmethod employs a \textbf{STR Predictor} to predict future GUIs' graphics and a \textbf{GUI-text Predictor} for generating the corresponding text. 
Moreover, we deploy \ourmethod 
to enhance agent-focused tasks by predicting the outcome of different action options.
Experiments show \ourmethod's ability to generate visually plausible and functionally effective GUIs that enable App agents to make more informed decisions. 
 Project link: \url{https://ai-agents-2030.github.io/ViMo/}

\end{abstract}

\section{Introduction}
Recent advancements in Large Language Models (LLMs)~\footnote{By LLMs, we refer to the concept of foundation models
that accept various input modalities (e.g., visual language
models (VLMs), multimodal LLMs (MLLMs)) while producing textual sequences \cite{wikipedia_llm}.} have unlocked new possibilities for deploying AI agents across diverse fields~\cite{li2023camel,goutora,rawles2024androidinthewild}. A notable application is the smartphone application~(App) agents~\cite{rawles2024androidworld,mobileagentv2}, designed to directly interact with Graphical User Interfaces (GUIs) to perform tasks autonomously and efficiently in a mobile operating system.

However, existing agents struggle with making decisions for tasks requiring longer steps \cite{web-world}. To address this "long-horizon" limitation, an increasing number of studies have introduced world models, which predict how GUIs evolve in response to user actions~\cite{isyour}. Yet, these models typically rely on language to describe future observations. 
These language-based descriptions often fail to capture the intricate visual details, such as the location and colour of GUI elements, necessary for a precise representation~\cite{web-world}. There is a need for a visual GUI world model capable of comprehensively representing observations in visual modality. Addressing this gap serves as the primary motivation for this work.

\begin{figure}[t]
    \centering
    \includegraphics[width=1\textwidth]{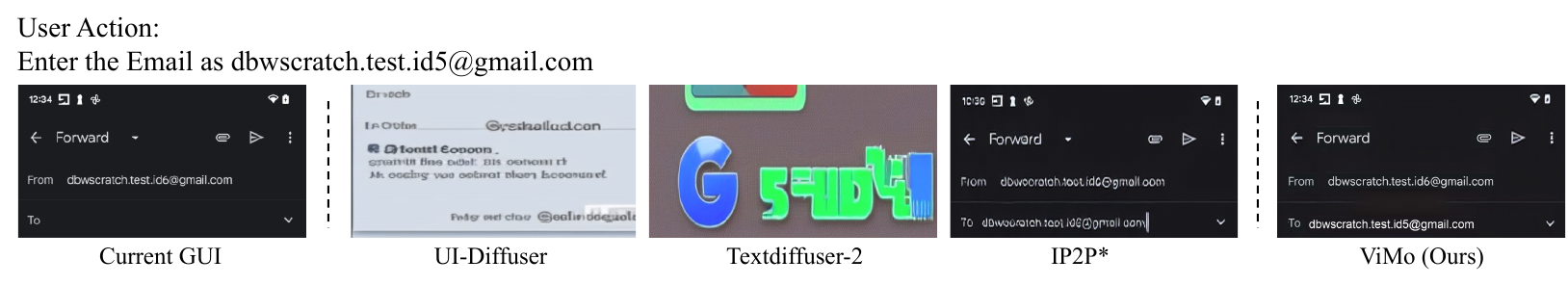} 
    \caption{ \dz{GUIs generated by image-based methods (UI-Diffuser~\cite{ai-diffuser}, TextDiffuser-2~\cite{text-diffuser2}, and IP2P~\cite{ip2p} fine-tuned on GUI dataset, denoted as IP2P*). Images are cropped for space efficiency.}}
    \label{fig:intro}
\end{figure}

To build a visual GUI world model capable of generating plausible future GUI observations that are visually consistent with user actions, a straightforward approach involves generating each pixel of a GUI using image generations ~\cite{ip2p,stablediffusion}. 
Although these methods demonstrate promising results, such as the GUI graphic generation on the location, style, and colour of GUI elements \cite{ai-diffuser}, or scene-text generation in a style that aligns the visual context \cite{text-diffuser2, scenetext}, they still 
display distortions in the text rendering, particularly for small-sized text where each pixel is critical for accurately identifying and representing the text~(see Fig. \ref{fig:intro} for an illustration).

To address the challenges of accurately generating high-fidelity text content within a GUI, 
we propose \textbf{\ourmethod}, the first visual GUI world model. \ourmethod decouples the generation of graphic and text content into distinct processes, using a novel data representation named Symbolic Text Representation (\textbf{STR}). In STR, each text content is replaced~(overlayed) with a text symbol, a rectangle-shaped placeholder with a defined border and fill colours, functioning as a special GUI element. 
Thus, we simplify the task of text content generation to text symbol generation, which reframes the problem to
the localisation of the text within a GUI. Based on STR, \ourmethod employs a \textbf{STR Predictor} and a \textbf{GUI-text Predictor} to generate the graphic and the text content respectively. Specifically, the STR predictor is implemented as a diffusion model, taking the current STR, extracted from the given GUI, and a user action as inputs to generate the STR of the next GUI. Meanwhile, the GUI-text predictor, implemented based on an LLM, leverages the STR generated by the STR predictor to produce the corresponding text for each text symbol. Finally, the predicted STR and the generated text are combined to produce the next GUI.

\dz{We evaluated \ourmethod in three distinct scenarios to comprehensively demonstrate its effectiveness.
First, we assessed its world model capability, where the quality of the
generated GUIs was measured using visual similarity, instructional
accuracy, and action readiness scores. Each score was examined through
both automatic metrics and user studies. These assessments provided a
robust and holistic evaluation of how visually precise and
contextually plausible the generated GUIs were. 
Second, we tested \ourmethod in an agent-focused task to evaluate
its benefits for existing App agents and its superiority over other
world models. In this setup, given a goal and the current App observation,
the agent selected optimal actions to achieve the goal
\cite{mobileagentv2}. By accurately predicting the next GUI based on
the current observation and an action, \ourmethod enabled the agent to better
anticipate action outcomes and make more informed decisions. This
experiment demonstrated the model’s effectiveness in enhancing
decision-making for App agents. 
Finally, we evaluated \ourmethod's real-world applicability under
two settings: online navigation and zero-shot generalisation. These
scenarios assessed the model's ability to perform in real-time
interactions and to generalise to previously unseen Apps, further
demonstrating its generalisation capabilities and practical value in
dynamic environments.}

Our main contributions are summarised as follows:

$\bullet$ We propose \ourmethod, the first generative visual GUI world model that predicts App observations in a visual modality, capable of more realistic and concrete visual GUI predictions compared to contemporary language-based methods.

$\bullet$ To address the challenge of strict pixel-level accuracy required to avoid distorted or blurred text generation in a GUI, we propose a Symbolic Text Representation (STR), overlaying text with uniform text symbols (placeholders) to simplify text content generation to text location generation. Then \ourmethod leverages an LLM to generate the corresponding text content for each text symbol.

$\bullet$  \dz{Extensive experiments demonstrated the effectiveness of \ourmethod in both world model evaluation and agent-focused tasks. Specifically, \ourmethod achieved an average 29.14\% and 182.74\% relative improvement over existing world models in terms of automatic metrics and user studies, respectively. Moreover, \ourmethod boosted the step-wise action prediction accuracy of App agents, achieving a 14.07\% relative performance gain.
In the online navigation setup, \ourmethod increased the task completion rate from 33.19\% to 40.95\%, yielding a substantial improvement of 7.76\%.}

\section{Related Works}

\subsection{App Agent} App agents, powered by LLMs, have made significant strides in automating tasks on mobile Apps~\cite{auto-droidv2, chen2025spa,zhang2024large,coat,mobile-gpt,nguyen2024gui}. These agents interact with GUIs by emulating human actions such as tapping, typing, and swiping. This enables them to accomplish various user goals, such as searching for products or setting alarms. Approaches in this domain are broadly divided into \textit{language-based} and \textit{multi-modality-based} methods. Language-based methods rely on textual description of the App observation and the user goal to generate appropriate actions~\cite{rawles2024androidworld,wen2024autodroid, android_control}, while multi-modality-based methods enhance this capability by incorporating GUIs for a more comprehensive understanding of the interface~\cite{rawles2024androidworld,mobileagentv2,christianos2025lightweight,wang2025distrl}. Despite the promising results, these approaches struggle with long-horizon tasks that require multiple interdependent actions and a deep understanding of dynamic environments~\cite{web-world}. For this challenge, recent works~\cite{agentq,koh2024tree} introduce the use of real-world emulators to simulate GUI changes resulting from agent actions, enabling App agents to better navigate complex, multi-step scenarios and improve decision-making accuracy. However, the use of real-world emulators faces significant drawbacks, including the high computational costs of setting up emulators for specific mobile systems and the safety risks involved in executing real-world interactions, such as repeatedly sending messages or making purchases. To overcome these, world models have attracted increasing attention as a more efficient alternative~\cite{web-world,isyour}.

\subsection{World Model }
By observing the real world, world models can predict how the environment evolves in response to an action \cite{lecun2022path,ding2024understanding}. For instance, video generation models such as Sora \cite{sora}  simulate future observations by predicting videos given a video observation and GameNGen \cite{gameeng} predicts how a game system will respond to user actions.  Notably, the ability to anticipate potential outcomes of actions has proven to be highly beneficial in informing decision-making processes~\cite{pascanu2017learning,yang2024evaluating,go,dreamer}. Inspired by their success, world models have emerged to predict the next observation on websites. These models~\cite{web-world,isyour,liu2023picture} typically take a website observation and an action as inputs to generate a textual description of the next observation. 
While websites provide multiple sources of information, including the actual site and their CSS or HTML source files, mobiles present a more limited context, as only the GUIs are typically accessible. 
Moreover, such text-only descriptions often lack the precise visual details required for accurately predicting future observations, highlighting the need for a visual world model capable of generating high-fidelity future GUI images.

\subsection{GUI Generation}
With the rapid advancements in image generation techniques, such as diffusion models~\cite{stablediffusion,customdiffusion,caoyu},
generating realistic images has become increasingly feasible. Building on these developments, previous methods have explored generating GUI directly in pixel space. For instance, layout generation methods generate the location of GUI elements \cite{layoutlu2023ui,layoutzheng2023layoutdiffusion,layoutsobolevsky2023guilget,layoutzhao2019image}, scene-text generation methods generate text that aligns with the visual context \cite{text-diffuser2,scenetext,text-diffuser,zeng2024textctrl}, UI-diffuser~\cite{ai-diffuser} fine-tune a stable diffusion model to generate mobile GUIs conditioned on text prompts. 
For the next GUI generation conditioned on current GUI observation and a user action, it seems straightforward to resort to an image-and-text-conditioned approach \cite{ip2p}. However, we find that pixel-based image generation struggles with rendering text accurately, as even minor pixel prediction errors can lead to distortions, particularly for small-sized text (see Fig.~\ref{fig:intro} for examples).

In this work, we advance beyond existing approaches that generate GUI entirely (graphic and text) at the pixel level. Instead, we render graphics as image pixels and generate text as language tokens, enabling a more accurate method for GUI generation.

\section{Method}
\label{sec:method}
In this section, we first define our setup in Subsection~\ref{subsec:Setup}. Then, we explain our motivation and introduce our \ourmethod in Subection~\ref{subsec:mvwm}. 
Finally, we demonstrate how \ourmethod can be applied to enhance existing App agents in real-world scenarios (Subsection~\ref{subsec:mvwm_agent}). All the prompts in this section are listed in the Appendix.

\subsection{Problem Setup}
\label{subsec:Setup}
In general, a GUI world model processes a given GUI Image $x_k$ at step $k$, and user action $a$, to predict the effect of $a$ on $x_k$ and simulate the next GUI. Formally, this can be expressed as:
\begin{equation}
x_{k+1}^a = f(x_k,a), 
\end{equation}

where $f(\cdot)$ represents the world model, and $x_{k+1}^a$ denotes the predicted next ($k$+1) GUI image after applying $a$ to $x_k$.  In the following, we explain in detail of our world model.

\begin{figure}[t]
    \centering
    \includegraphics[width=1\textwidth]{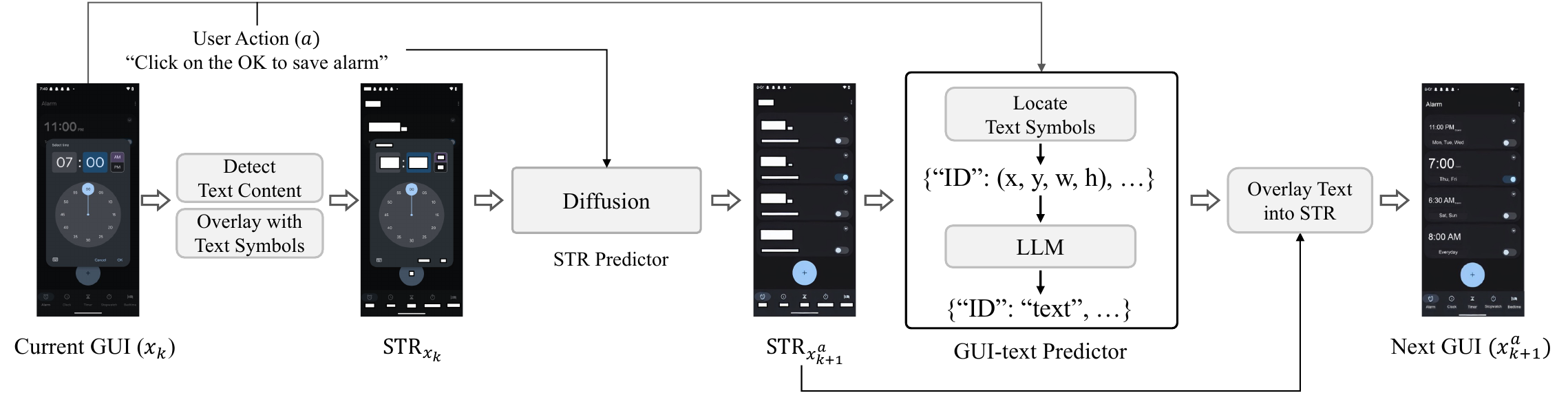} 
    \caption{Framework of our \ourmethod. We first detect text content (actual words) in the current GUI ($x_k$) and overlay it with text symbols~(rectangle-shaped placeholders with a black border and white fill),  to create $\text{STR}_{x_k}$. Then $\text{STR}_{x_k}$ and the user action ($a$) are input to the STR predictor to generate the STR of the next GUI ($\text{STR}_{{x^a_{k+1}}}$). Next, text symbols within $\text{STR}_{{x^a_{k+1}}}$ are located and assigned unique ID token. Then the LLM predicts the text content corresponding to each token. Finally, the next GUI image is constructed by overlaying the predicted text into the STR.}
    \label{fig:method_1}
\end{figure}

\subsection{\ourmethod: Generative Visual GUI World Model}
\label{subsec:mvwm}

To tackle the limitation of existing methods \cite{ai-diffuser,text-diffuser2,ip2p} in generating visually plausible text for a GUI, as shown in Fig. \ref{fig:intro}, we propose \ourmethod, a novel generative visual GUI world model that decouples the graphic and text content generation. As shown in Fig. \ref{fig:method_1}, we first detect and remove all the text in the GUI by overlaying it with a text symbol to create the Symbolic Text Representation (STR).  Then a STR predictor is leveraged for determining the STR representation of the next GUI with a pixel-based diffusion process. Finally, a GUI-text predictor is proposed to generate the text content for each symbol using an LLM, followed by a handcrafted design to overplay the text into the STR image to create the next GUI. Their details are specified in the following.

\subsubsection{STR: Symbolic Text Representation}
\label{subsubsection:STR}
To develop a GUI prediction model that eliminates the need to generate specific text content, we propose the Symbolic Text  Representation (STR), where all the text content (actual words) within the GUI image is symbolised (overlayed) with uniform text symbols (placeholders).  To be specific, we create an STR representation from a given GUI image with three steps: 1) using an OCR model~\cite{paddleocr,qiao2020seed} to detect text within the GUI; 2) masking the detected text by overlaying it with a box filled with white and bordered in black; 3) we leverage an LLM to filter out static text displayed on static GUI elements and preserve it in the image, as it does not involve any semantic evolution or dynamic changes and remains unchanged as part of specific elements such as a keyboard or a clock face.  Additionally, we empirically find that predicting this static text with complex spatial patterns poses significant challenges for the LLM.

Through the above process, GUI images are transformed into the STR representation, where the text content is abstracted into a text symbol, relaxing the task of generating semantic text content into predicting text symbols that indicate the location and size.

\subsubsection{STR Predictor}
Building on the powerful generative capability of diffusion-based models~\cite{stablediffusion}, we introduce a STR predictor specifically trained to understand a given STR and a user action, enabling it to generate the corresponding next STR effectively. In particular, we fine-tune a pre-trained stable diffusion model \cite{stablediffusion} to predict the next STR, conditioned on the STR of the current GUI and the user action. Given a STR representation~($\text{STR}_{x_{k}}$) extracted from GUI~($x_k$),
the process starts with the encoding of $\text{STR}_{x_{k}}$into a latent representation \cite{vae}: $z = \mathcal{E}(\text{STR}_{x_{k}})$.
Gaussian noise is then added to this representation to create \( z_t \) at timestep \( t \). A denoising autoencoder is subsequently trained to predict the Gaussian noise in the latent representation, aiming to reverse the noise addition. The objective is defined as:
\begin{equation}
\setlength{\abovedisplayskip}{2pt}
\setlength{\belowdisplayskip}{2pt}
L= \mathbb{E}_{\mathcal{E}(\text{STR}_x), \epsilon \sim \mathcal{N}(0, I), t}\Big[\| \epsilon - \epsilon_\theta(z_{t},\mathcal{E}(\text{STR}_{x_{k}}), t, a) \|_{2}^{2}\Big],
\end{equation}
where \( \epsilon_\theta \) is a U-Net \cite{unet} architecture 
conditioned on a timestep $t$, a text prompt  $a$ (action), the visual input $z_t$ and the image condition $\text{STR}_{x_{k}}$. To support the condition on images, we follow IP2P \cite{ip2p} to add additional input channels to the first convolutional layer, concatenating the image condition~$\mathcal{E}(\text{STR}_{x_{k}})$ with the noised latent $z_t$. After training, our STR predictor is capable of synthesising the next STR ($\text{STR}_{x^a_{k+1}}$) for given  $\text{STR}_{x_{k}}$  with action instruction $a$.

\subsubsection{GUI-Text Predictor}
Given a STR representation generated by our STR predictor, we design a GUI-text predictor to generate plausible text for the text symbols in the STR based on its graphics. Specifically, we first locate the text symbols in the STR by colour matching and boundary detection. This outputs the location of text symbols, along with their unique ID tokens assigned via enumeration, denoted as $\mathcal{T}$. Then we leverage the image processing and task understanding ability of LLM to predict the text content based on its context in STR. This process can be formulated as:
 \begin{equation}
\setlength{\abovedisplayskip}{2pt}
\setlength{\belowdisplayskip}{2pt}
\text{text}_{x^a_{k+1}} = \mathrm{LLM}(\text{STR}_{x^a_{k+1}},x_k,a,\mathcal{T}),
\end{equation}
where $\text{STR}_{x^a_{k+1}}$ denotes the STR representation of $x^a_{k+1}$. $\text{text}_{x^a_{k+1}}$ contains the predicted text content for each text symbol, associated with its ID token. This design ensures flexible and accurate text generation tailored to the predicted GUI STRs as the context. Finally, we overlay each text content~($\text{text}_{x^a_{k+1}}$) to $\text{STR}_{x^a_{k+1}}$ to reconstruct the predicted GUI image ($x^a_{k+1}$). To be specific, text symbols are replaced with the corresponding text based on coordinates, with dynamic styling determined by the symbol's size and surrounding colours. More details are provided in the Appendix.

\begin{algorithm}[t]
   \caption{Enhancing App Agent with Generative Visual GUI World Model}
   \label{alg}
\begin{algorithmic}
   \STATE {\bfseries Input:}  Current GUI Observation $x_k$,
  A goal $g$,
   A visual world model \ourmethod,
   A selection model $S(\cdot)$. 
   \STATE {\bfseries Output:}   Action to be applied on $x_k$ to achieve  $g$. 
   
   Generate action options $\mathcal{A}$ with  $n$ actions {$\{a^i\}$} (Eq. ~\eqref{eq:3}).
   \FOR{$i=1$ {\bfseries to} $n$}
\STATE Leverage \ourmethod to synthesis the next GUI observation conditioned on $a^i$ and $x_k$, denoted as $x_{k+1}^{a^i}$~(Eq.~\eqref{eq:4}).
   \ENDFOR
   
Use $S(\cdot)$ to identify the optimal action with predicted observation (Eq.~(\ref{eq:5})).
\end{algorithmic}
\end{algorithm}

\subsection{\ourmethod Enhanced App Agent}
\label{subsec:mvwm_agent}
Motivated by that App agents usually face limitations in long-horizon planning to make optimal decisions on action selection \cite{web-world}, we leverage the proposed \ourmethod to enhance the decision-making of App agents.

To be specific, we break down the process into three steps: action option generation, action outcome synthesis, and action selection.
In the first step, the App agent generates $n$ action options, as follows:
\begin{equation}
\label{eq:3}
\setlength{\abovedisplayskip}{2pt}
\setlength{\belowdisplayskip}{2pt}
\mathcal{A} =\text{Agent}(x_k, g),
\end{equation}
where $\mathcal{A} = \{a^{1}, a^{2}, \cdots, a^{n}\}$ denotes the action option set, $x_k$ is the current GUI image at step $k$, and $g$ is the given user goal. With these action options, our world model \ourmethod is leveraged to synthesise the outcome (next GUI) of these actions as follows:
\begin{equation}
\label{eq:4}
\setlength{\abovedisplayskip}{2pt}
\setlength{\belowdisplayskip}{2pt}
x_{k+1}^{a^i} =\textbf{\ourmethod}(x_k,a^{i}),
\end{equation}

where $x_{k+1}^{a^i}$ denotes the synthesised next GUI of applying action $a^i$ on $x_k$. Finally, each action $a^i$ and its corresponding predicted outcome $x_{k+1}^{a^i}$ are fed into an LLM-based selection model, which identifies the optimal action based on the generated GUIs. This process can be formulated as:
\begin{equation}
\label{eq:5}
\setlength{\abovedisplayskip}{2pt}
\setlength{\belowdisplayskip}{2pt}
a_{se} = S\left(\{(a^i, x_{k+1}^{a^i})\}_{i=1}^{n}\right),
\end{equation}
where $a_{se}$ denotes the selected action, and $S(\cdot)$ is the selection model. This procedure is outlined in Algorithm \ref{alg}. By predicting the next GUI, \ourmethod provides the agent with the potential outcome of an action, enabling it to make more informed decisions. 
\dz{We build the selection model to identify the best action in
  two steps. First, we query an LLM to evaluate all the action
  candidates, providing a judgment—either \textit{valid} or
  \textit{invalid}—and a confidence score for each action. Second, we
  query the LLM again to select the best action from the two
  highest-scoring actions. This process is motivated by our
  observation that, in over $70\%$ of tasks, the difference between
  the top two scores is equal to or less than $0.1$, indicating that
  both are likely optimal.  By explicitly prompting the LLM to compare
  the top candidates, we go beyond coarse scoring and enable more
  detailed decision-making.}

\begin{table}[t] 
    \centering
      \setlength{\tabcolsep}{0.18cm}

    \caption{GUI quality evaluation. $s_{gc}$ indicates the GUI consistency, $s_{ia}$  instructional accuracy and $s_{ar}$ action readiness score. $s_{h}$ indicates the harmonic average between the 3 metrics. $\Delta s_h$ is the relative performance gains of our \ourmethod over other methods. \dz{IP2P* denotes finetuing of IP2P on our dataset.} 
    }
    \begin{tabular}{l|ccccc|ccccc}
        \toprule
        \multirow{2}{*}{Method}
 &\multicolumn{5}{c|}{Automatic Metric} &\multicolumn{5}{c}{\dz{User Study}}
        \\
        \cline{2-11}
        &$s_{gc}$ &$s_{ia}$ &$s_{ar}$&$s_{h}$ &$\Delta s_h$&$s_{gc}$ &$s_{ia}$ &$s_{ar}$&$s_{h}$ &$\Delta s_h$
\\
        \hline
              HTML-vision &0.70&\textbf{85.77}&62.79&0.72&5.39\% &0.31&11.32&9.01&0.23&282.61\%\\
      IP2P*&\textbf{0.74}&63.57&70.15 &0.69&10.20\%&0.82&58.92&52.81 &0.63&39.68\%\\
      UI-diffuser &0.60&39.61&38.75&0.44&71.82\% &0.36&14.32&8.56 &0.27&225.93\%\\
\hline
      \ourmethod (Ours)&\textbf{0.74} &75.39&\textbf{78.68}&\textbf{0.76} &-&\textbf{0.89}&\textbf{91.12}&\textbf{84.71}&\textbf{0.88}&-\\
        \bottomrule
        
    \end{tabular}
        \label{tab:quanti_com}
\end{table}

\section{Experiments}

In this section, we begin by summarising our proposed STR dataset
discussed in Subsection \ref{subsubsection:STR}. Next, we tested the core capability of our \ourmethod, focusing on its GUI generation
ability. Building on this, we demonstrated how the powerful GUI
generation capability of \ourmethod can enhance the decision-making of
App agents. \dz{Then, we studied our effectiveness in real-world
  App navigation tasks.} Finally, we carried out the ablation study to
validate the effectiveness of our model design. Specific setups and
experiment details are elaborated in subsequent sections. \dz{ Unless
  explicitly stated otherwise, GPT-4o \cite{gpt4o}  was employed as the default LLM
  in the following sections.}

 \subsection{Dataset Summarisation}
\label{sub:datasum}
Our STR dataset was constructed using data from two widely recognised and large-scale sources: Android Control~\cite{android_control} and Android in the Wild~(AITW)~\cite{rawles2024androidinthewild}. From these sources, we respectively sampled 12 and 7 Apps, selecting those with rich data samples while filtering out noise.
Android Control provides two types of user actions: 1) \textit{action commands}: predefined actions (e.g., click, scroll) accompanied by specific parameters such as coordinates (x, y); 2) \textit{action instructions}: actions described in natural languages, such as "click the plus icon".
We used action instructions as conditions for our world model as this approach was more concrete and better utilised the pre-trained model in understanding natural language.
For AITW, action commands were converted into action instructions using GPT-4o \cite{gpt4o}. 
\dz{In total, we collected 19 Apps with 3,550 episodes, 23,620 images and 18,450 actions.}
To ensure both time-efficient and cost-efficient experiments, we followed prior App agent~\cite{rawles2024androidinthewild} on partial split evaluation, instead of evaluating the model on the entire testing set. Specifically, we randomly sampled 57 episodes across 19 distinct Apps.
Details including dataset collection, split summarisation and full-split experiments are provided in the Appendix.

\begin{figure*}[t]
    \centering
    \includegraphics[width=0.95\textwidth]{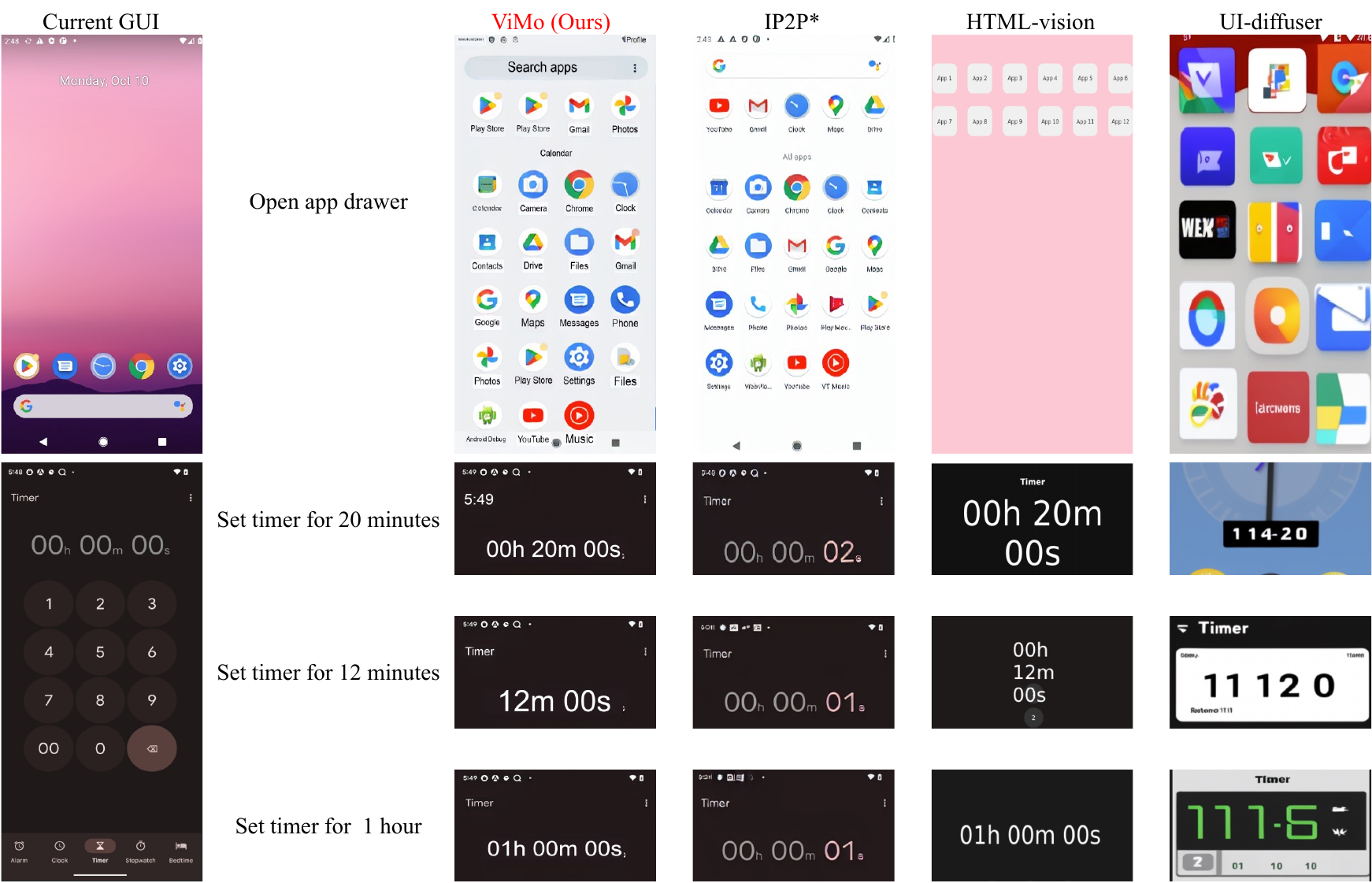} 
    \caption{GUI generation comparison in graphic generation (Top) and text generation (Bottom).
    }
    \label{fig:vis_compare}
\end{figure*}

\subsection{World Model Ability}
\label{subsec: wma}
\dz{We evaluated the GUI generation capability of \ourmethod by GUI quality evaluation. We included IP2P~\cite{ip2p} and UI-diffuser \cite{ai-diffuser}, both originally designed for image editing and GUI generation. For IP2P, we fine-tuned their diffusion model on our dataset to generate everything of the GUI, including the text content and the graphic, denoted as IP2P*. We also leveraged an LLM to predict App observations in an HTML format, which were rendered into images, denoted as HTML-vision.}

\dz{We leveraged 3 evaluation metrics: The \textit{GUI consistency score ($s_{gc}$)} assessed the visual similarity between the ground truth and the generated next GUI; \textit{Instructional accuracy score ($s_{ia}$)} determined whether the generated GUI adheres to the user action;  \textit{Action readiness score ($s_{ar}$}) evaluated whether the generated GUI retains valid elements essential for subsequent actions required to achieve the user goal. 
We conducted both automatic evaluations and user preference studies to assess our methods. For the automatic evaluation, we used DINO \cite{DINO} as the visual encoder to compute $s_{gc}$, and an LLM to evaluate $s_{ia}$ and $s_{ar}$. The prompts used for the LLM-based evaluations are provided in the Appendix.
For the user study, we invited 70 voluntary participants to
complete questionnaires based on 80 GUI samples,
generated by all 4 compared methods. For each sample,
participants were asked three questions—one each for evaluating
$s_{gc}$, $s_{ia}$, and $s_{ar}$.  The full instructions provided to the participants are provided in the Appendix.}

As shown in Table~\ref{tab:quanti_com}, \ourmethod achieved the highest score on the harmonic average of the three automatic metrics, surpassing other methods with an average relative performance improvement of 29.14\%. \dz{ The results of the user study were consistent with the automatic evaluations, where \ourmethod demonstrated the best performance. Notably, HTML-vision and UI-diffuser performed significantly worse in human evaluation compared to their scores in the LLM-based assessment. This discrepancy likely arose because human evaluators perceived the outputs of these methods as visually unrealistic or functionally incoherent, leading to lower subjective scores in $s_{gc}$, $s_{ia}$ and $s_{h}$.}

Qualitative comparisons are presented in Fig.~\ref{fig:vis_compare}, under two scenarios: GUI graphic changes (Top) and text generation (Bottom, cropped for space efficiency). Experiments revealed that while the HTML-vision method exhibited greater flexibility in responding to user actions (as shown in the bottom examples), it failed to produce concrete details necessary for future actions (top). Conversely, IP2P* generated plausible GUI graphics but lacked flexibility in text content generation (also reflected by $s_{gc}$ and $s_{ia}$ in Table~\ref{tab:quanti_com}). This trade-off highlighted the superior balance of \ourmethod.

\subsection{World Model Enhanced App Agent}

\begin{table}[!t]
\centering
\caption{Decision optimisation comparison with App agents.
}
\setlength{\tabcolsep}{4mm}

\resizebox{1\columnwidth}{!}{

\begin{tabular}{l|lcccc}
\toprule
Agent Type &App Agent& Leisure & Work & System &Overall \\
\hline
\multirow{4}{*}{Language-Based} & ER \cite{android_control} &31.76 & 46.15& 34.13&34.50 \\
&AutoDroid \cite{wen2024autodroid}&35.81 &46.15&31.75&35.46\\
&T3A \cite{rawles2024androidworld}&41.22&51.28&42.86&43.13\\
\cline{2-6}
&T3A + \ourmethod (\textbf{Ours})& 50.00&\textbf{58.97}&\textbf{45.24}&49.20\\
\hline
\multirow{4}{*}{Multi-Modality-Based} &APP-Agent \cite{zhang2023appagent} & 43.24&51.28&39.68&42.81\\
&Mobile-Agent-v2 \cite{mobileagentv2} & 43.92&53.85&39.68&43.45\\
&M3A \cite{rawles2024androidworld} & 46.62&51.28&43.65& 46.01\\
\cline{2-6}
&M3A + \ourmethod (\textbf{Ours})&\textbf{53.38}&53.85&\textbf{45.24}&\textbf{50.16}\\
\bottomrule
\end{tabular}
}
\label{table:agent_com}
\end{table}

This section demonstrates that: 1) \ourmethod enhanced the performance of App agents in decision-making; 2) \ourmethod outperformed other world models in enabling App agents to make more accurate decisions.

\textbf{Comparison with App Agents.} In this experiment, we collected 6 LLM-based App agents, which included three language-based methods: ER, AutoDroid, and T3A, as well as three multi-modality-based methods: APP-Agent, Mobile-Agent-v2, and M3A. We applied our \ourmethod into M3A and T3A following the process in Subsection \ref{subsec:mvwm_agent}. Moreover, we followed the previous works~\cite{rawles2024androidinthewild,android_control} to use the step accuracy (the number of correct actions divided by the number of overall actions) to quantify the model performance. \dz{To provide more detailed results, we categorised the Apps into three groups: "Leisure", "Work" and "System".} Table~\ref{table:agent_com} demonstrates that \ourmethod was beneficial to the App agent, achieving a relative performance gain of 9.01\% for M3A and 14.07\% for T3A. These findings highlighted the effectiveness of our proposed world model in providing App agents with enhanced decision-making capability. Additional information about the categorisation and experiments with more App agents are provided in the Appendix.

\textbf{Comparison with World Models.} We further evaluated the ability of \ourmethod to enhance App agent decision-making by comparing it against \dz{existing world models.  In addition to vision-based world models discussed in Subsection \ref{subsec: wma}, we also incorporated two language-based world models where we adapted website world models \cite{isyour} to App setups, utilising Change-text to generate textual descriptions capturing differences between consecutive observations and HTML-text to predict App observations in an HTML format.}\dz{ Then, we applied all the world models on M3A App agents.} 
 Table \ref{table:compare_wm} illustrates that our proposed world model outperformed other world models, achieving a step accuracy (Step Acc.) of 50.16\%. These findings underscored the effectiveness of our world model in enhancing step-level decision-making. Experiments with more App agents are provided in the Appendix.

\begin{table}[t]
\vspace{-10pt}
    \centering
    \scriptsize
      \begin{minipage}[t]{0.32\textwidth}
         \centering
                 \setlength{\tabcolsep}{5pt}

         \caption{Result on M3A Agent.}
         \label{table:compare_wm}

             \begin{tabular}{l|c|c}
        \toprule
 Modality&World Model & Step Acc.\\
\hline

w/o WM&w/o WM&46.01\\
\hline
\multirow{2}{*}{Language} &Change-text&47.28\\
& HTML-text&46.65\\
\hline
\multirow{4}{*}{Vision}&HTML-vision&48.89 \\
&UI-diffuser&47.60\\
&IP2P*&48.56\\
\cline{2-3}
&\textbf{\ourmethod} (\textbf{Ours})&\textbf{50.16}\\
\hline
\end{tabular}
        \end{minipage}
        \hfill
    \begin{minipage}[t]{0.32\textwidth}
        \centering
\captionof{table}{\dz{Zero-shot Evaluation.}}
\label{table:zerosshot}
        \setlength{\tabcolsep}{2pt}
\begin{tabular}{l|c|c}
\toprule
App Agent &LLM& Step Acc. \\
\hline
SeeAct&GPT-4-Turbo & 33.9 \\
M3A&GPT-4-Turbo & 42.1 \\
ER&Gemini 1.5 Pro & 24.4 \\
\hline
T3A&Gemini-2.0-Flash& 41.4\\
T3A+\ourmethod&Gemini-2.0-Flash& 46.8 \\
\hline
M3A&Gemini-2.0-Flash& 44.2\\ 
M3A+\ourmethod&Gemini-2.0-Flash& \textbf{47.6} \\
            \hline
        \end{tabular}
    \end{minipage}
\hfill
      \begin{minipage}[t]{0.33\textwidth}

\centering
\caption{\dz{Online Evaluation.}
}        \label{table:online_test}
        \setlength{\tabcolsep}{2pt}

\begin{tabular}{l|c|c}
\toprule
App Agent & 
LLM
& Task Acc.  \\
\hline
SeeAct & GPT-4-Turbo & 15.50 \\
M3A & GPT-4-Turbo & 25.40 \\
M3A &  Gemini-1.5-Pro &  22.80 \\
T3A & GPT-4-Turbo& 30.60 \\
T3A & Gemini-1.5-Pro & 19.40 \\
T3A & Gemini-2.0-Flash  & 33.19 \\
\hline
T3A + \ourmethod & Gemini-2.0-Flash  & \textbf{40.95} \\
\hline
\end{tabular}
\end{minipage}

\end{table}

\subsection{\dz{Real-world Applications}}

\dz{

\paragraph{
Practical Deployment.} \ourmethod was designed to be lightweight and easily deployable. The minimum requirement for deployment is a GPU with 16 GB of memory. Moreover, \ourmethod was implemented as a plug-and-play API that required only a single function call, making integration straightforward.  Inference time on V100 GPU is 8 seconds on a STR image generation and 30 seconds on GUI-text prediction.  We collected and compared the inference time with existing methods in the Appendix.

\paragraph{Generalisation to New Apps.} \dz{Generalisation is a crucial capability for real-world applications. To assess the generalisation performance of our method on new Apps that were unseen during training, we conducted a zero-shot evaluation using data from the Android Control dataset~\cite{android_control}, explicitly excluding Apps encountered during training. As shown in Table~\ref{table:zerosshot}, our approach substantially outperformed the baseline and achieved 47.6\%, underscoring its robustness and adaptability to novel App environments. Additional visualisations of unseen scenarios are provided in the Appendix.}

\paragraph{Online Evaluation.} To further demonstrate the effectiveness of \ourmethod in realistic App navigation scenarios, we conducted an online evaluation using the AndroidWorld dataset~\cite{rawles2024androidworld}, which comprises 116 distinct navigation tasks. Performance was measured using the task success rate (Task Acc.). As illustrated in Table~\ref{table:online_test}, \ourmethod achieved a notable improvement of 7.76\% over the baseline method, highlighting its effectiveness and reliability in real-world settings.

}

\subsection{Ablation Study}
\label{subsec: ablation}
\begin{wrapfigure}{r}{0.45\linewidth}          
  \vspace{-1.0\baselineskip}                   
  \centering
  \footnotesize                                
  \setlength{\tabcolsep}{4pt}
  \renewcommand{\arraystretch}{1.12}

  \begin{minipage}{\linewidth}
    \captionof{table}{Ablations on preserving static text and using action instructions. }
    \label{tab:ablation_two}
    \vspace{-.5em}
    \centering
    \begin{tabular}{@{}cc|cc@{}}
      \toprule
      \multirow{2}{*}{Static Text} & \multirow{2}{*}{Action Instr.} & \multicolumn{2}{c}{App Agent} \\
      \cmidrule(l){3-4}
       & & T3A & M3A \\ \midrule
       N/A        &  N/A           & 43.13 & 46.01 \\
       \hline
       \checkmark & --          & 42.81 & 45.05 \\
       --         & \checkmark  & 47.28 & 48.88 \\
       \checkmark & \checkmark  & \textbf{49.20} & \textbf{50.16} \\
      \bottomrule
    \end{tabular}
  \end{minipage}
  \vspace{0.9\baselineskip}                    

  \begin{minipage}{\linewidth}
    \captionof{table}{Ablation on the number of iterations.}
    \label{table:steps}
    \vspace{-.5em}
    \centering
    \begin{tabular}{@{}lcc@{}}
      \toprule
      Method & Iterations & Step Acc.\ (\%) \\ \midrule
      T3A                       & N/A & 39.94 \\ \cmidrule(l){1-3}
      \multirow{3}{*}{T3A+\ourmethod}
          & 1 & 46.06 \\
          & 2 & \textbf{46.65} \\
          & 3 & 45.05 \\
      \bottomrule
    \end{tabular}
  \end{minipage}
\vspace{-2em}
\end{wrapfigure}

In this section, we ablated on three key components of \ourmethod: 1) preserving static text within the image to simplify the text generation task; 2) using action instructions instead of action commands as the conditioning input for \ourmethod; and \dz{3) varying the number of iterations, where each iteration corresponds to one roll-out step into the future during GUI prediction. 
}

Firstly, for the challenge of predicting static text from specific GUI elements, such as keyboard, number pad or clock face, which typically did not involve text changes and exhibited complex spatial patterns, we retained static text within the image (Subsection \ref{subsubsection:STR}). This approach eliminated the need for the LLM to generate such static text while generating in pixels instead. 
Secondly, we proposed conditioning STR prediction on action instruction rather than action commands (Subsection \ref{sub:datasum}).
\dz{Ablation results are
presented in Table~\ref{tab:ablation_two}, where "Static Text" indicates whether static text was retained in the images, and "Action Instr." denotes whether natural language instructions ("$\surd$") or abstract action commands~("-") were used as conditioning input to \ourmethod. The first row indicates the baseline where \ourmethod was not applied. The table shows that both components contributed significantly to performance improvements across the two App agents, highlighting their critical roles in enabling \ourmethod to generate high-quality GUIs. Visual comparison examples are provided in the Appendix for further illustration.}

\dz{Our \ourmethod predicted future GUI observations, which could be
  recursively fed back as input to simulate further into the
  future. In this ablation study, we varied the iteration number
  to evaluate how extended roll-outs impact prediction accuracy. We
  took Gemini-2.0-Flash \cite{gemini2024} as the LLM in this
  study.  As shown in Table\ref{table:steps}, performing two
  iterations yielded the highest accuracy. However, this also led to
  increased computational cost.  Therefore, we selected one step as a
  practical trade-off between performance and efficiency. We also
  observed a slight decline in performance at iteration 3 relative to
  iterations 1 and 2, indicating that extending the prediction horizon
  did not necessarily improve agent behaviour. This was likely due
    to that longer horizons introduced not only additional foresight
  but also a greater accumulation of prediction errors, whose
  detrimental effect could outweigh the potential benefits. Further
  analysis of error accumulation across iterations, along with
  comparisons with various world models and implementation details, is provided
  in the Appendix.}

\section{Conclusion}
In this work, we introduced \ourmethod, a novel generative visual GUI world model designed to predict App observations in a visual modality, providing a more realistic and concrete approach compared to contemporary language-based models. To address the unique challenges of GUI generation, \ourmethod was equipped with the Symbolic Text Representation~(STR) to simplify text content generation to text location prediction by overlaying text content with placeholders and delegating content generation to LLM. This innovation ensured high visual fidelity and avoided artefacts like distorted or blurred text. Through extensive experiments, we demonstrated that \ourmethod generated both visually plausible and functionally effective GUIs.
Notably, \ourmethod boosted step-wise action prediction accuracy by a relative performance gain of 14.07\%, underscoring its potential to enhance decision-making  of App agents. \dz{Furthermore, real-world experiments demonstrated the strong generalisation ability of \ourmethod to unseen Apps, along with its robust performance in online navigation tasks under real-time environment interaction.}
These results highlighted the value of \ourmethod for advancing GUI world models and set a new benchmark for future
research.

{
  \small
  \bibliographystyle{unsrt}
  \bibliography{ref}\label{reference}

\begin{thebibliography}{10}

\bibitem{wikipedia_llm}
{W. contributors}.
\newblock {Large language model --- Wikipedia, the free encyclopedia}.
\newblock [Online]. Available:
  \\url{https://en.wikipedia.org/wiki/Large\_language\_model}, 2024.
\newblock Accessed: 2024-11-25.

\bibitem{li2023camel}
Guohao Li, Hasan Hammoud, Hani Itani, Dmitrii Khizbullin, and Bernard Ghanem.
\newblock Camel: Communicative agents for" mind" exploration of large language
  model society.
\newblock {\em NeurIPS}, 36:51991--52008, 2023.

\bibitem{goutora}
Zhibin Gou, Zhihong Shao, Yeyun Gong, Yujiu Yang, Minlie Huang, Nan Duan,
  Weizhu Chen, et~al.
\newblock Tora: A tool-integrated reasoning agent for mathematical problem
  solving.
\newblock In {\em ICLR}, 2023.

\bibitem{rawles2024androidinthewild}
Christopher Rawles, Alice Li, Daniel Rodriguez, Oriana Riva, and Timothy
  Lillicrap.
\newblock Androidinthewild: A large-scale dataset for android device control.
\newblock {\em NeurIPS}, 36, 2024.

\bibitem{rawles2024androidworld}
Christopher Rawles, Sarah Clinckemaillie, Yifan Chang, Jonathan Waltz,
  Gabrielle Lau, Marybeth Fair, Alice Li, William Bishop, Wei Li, Folawiyo
  Campbell-Ajala, et~al.
\newblock Androidworld: A dynamic benchmarking environment for autonomous
  agents.
\newblock {\em arXiv preprint arXiv:2405.14573}, 2024.

\bibitem{mobileagentv2}
Junyang Wang, Haiyang Xu, Haitao Jia, Xi~Zhang, Ming Yan, Weizhou Shen,
  Ji~Zhang, Fei Huang, and Jitao Sang.
\newblock Mobile-agent-v2: Mobile device operation assistant with effective
  navigation via multi-agent collaboration.
\newblock {\em arXiv preprint arXiv:2406.01014}, 2024.

\bibitem{web-world}
Hyungjoo Chae, Namyoung Kim, Kai Tzu-iunn Ong, Minju Gwak, Gwanwoo Song, Jihoon
  Kim, Sunghwan Kim, Dongha Lee, and Jinyoung Yeo.
\newblock Web agents with world models: Learning and leveraging environment
  dynamics in web navigation.
\newblock {\em arXiv preprint arXiv:2410.13232}, 2024.

\bibitem{isyour}
Yu~Gu, Boyuan Zheng, Boyu Gou, Kai Zhang, Cheng Chang, Sanjari Srivastava,
  Yanan Xie, Peng Qi, Huan Sun, and Yu~Su.
\newblock Is your llm secretly a world model of the internet? model-based
  planning for web agents.
\newblock {\em arXiv preprint arXiv:2411.06559}, 2024.

\bibitem{ai-diffuser}
Jialiang Wei, Anne-Lise Courbis, Thomas Lambolais, G{\'e}rard Dray, and Walid
  Maalej.
\newblock On ai-inspired ui-design.
\newblock {\em arXiv preprint arXiv:2406.13631}, 2024.

\bibitem{text-diffuser2}
Jingye Chen, Yupan Huang, Tengchao Lv, Lei Cui, Qifeng Chen, and Furu Wei.
\newblock Textdiffuser-2: Unleashing the power of language models for text
  rendering.
\newblock In {\em ECCV}, pages 386--402. Springer, 2024.

\bibitem{ip2p}
Tim Brooks, Aleksander Holynski, and Alexei~A Efros.
\newblock Instructpix2pix: Learning to follow image editing instructions.
\newblock In {\em CVPR}, pages 18392--18402, 2023.

\bibitem{stablediffusion}
Robin Rombach, Andreas Blattmann, Dominik Lorenz, Patrick Esser, and Bj{\"o}rn
  Ommer.
\newblock High-resolution image synthesis with latent diffusion models.
\newblock In {\em CVPR}, pages 10684--10695, 2022.

\bibitem{scenetext}
Lingjun Zhang, Xinyuan Chen, Yaohui Wang, Yue Lu, and Yu~Qiao.
\newblock Brush your text: Synthesize any scene text on images via diffusion
  model.
\newblock In {\em AAAI}, volume~38, pages 7215--7223, 2024.

\bibitem{auto-droidv2}
Hao Wen, Shizuo Tian, Borislav Pavlov, Wenjie Du, Yixuan Li, Ge~Chang, Shanhui
  Zhao, Jiacheng Liu, Yunxin Liu, Ya-Qin Zhang, et~al.
\newblock Autodroid-v2: Boosting slm-based gui agents via code generation.
\newblock {\em arXiv preprint arXiv:2412.18116}, 2024.

\bibitem{chen2025spa}
Jingxuan Chen, Derek Yuen, Bin Xie, Yuhao Yang, Gongwei Chen, Zhihao Wu,
  Li~Yixing, Xurui Zhou, Weiwen Liu, Shuai Wang, et~al.
\newblock Spa-bench: A comprehensive benchmark for smartphone agent evaluation.
\newblock In {\em NeurIPS 2024 Workshop on Open-World Agents}, 2024.

\bibitem{zhang2024large}
Chaoyun Zhang, Shilin He, Jiaxu Qian, Bowen Li, Liqun Li, Si~Qin, Yu~Kang,
  Minghua Ma, Qingwei Lin, Saravan Rajmohan, et~al.
\newblock Large language model-brained gui agents: A survey.
\newblock {\em arXiv preprint arXiv:2411.18279}, 2024.

\bibitem{coat}
Jiwen Zhang, Jihao Wu, Yihua Teng, Minghui Liao, Nuo Xu, Xiao Xiao, Zhongyu
  Wei, and Duyu Tang.
\newblock Android in the zoo: Chain-of-action-thought for gui agents.
\newblock {\em arXiv preprint arXiv:2403.02713}, 2024.

\bibitem{mobile-gpt}
Sunjae Lee, Junyoung Choi, Jungjae Lee, Munim~Hasan Wasi, Hojun Choi, Steven~Y
  Ko, Sangeun Oh, and Insik Shin.
\newblock Explore, select, derive, and recall: Augmenting llm with human-like
  memory for mobile task automation.
\newblock {\em arXiv preprint arXiv:2312.03003}, 2023.

\bibitem{nguyen2024gui}
Dang Nguyen, Jian Chen, Yu~Wang, Gang Wu, Namyong Park, Zhengmian Hu, Hanjia
  Lyu, Junda Wu, Ryan Aponte, Yu~Xia, et~al.
\newblock Gui agents: A survey.
\newblock {\em arXiv preprint arXiv:2412.13501}, 2024.

\bibitem{wen2024autodroid}
Hao Wen, Yuanchun Li, Guohong Liu, Shanhui Zhao, Tao Yu, Toby Jia-Jun Li, Shiqi
  Jiang, Yunhao Liu, Yaqin Zhang, and Yunxin Liu.
\newblock Autodroid: Llm-powered task automation in android.
\newblock In {\em MobiCom}, pages 543--557, 2024.

\bibitem{android_control}
Wei Li, William Bishop, Alice Li, Chris Rawles, Folawiyo Campbell-Ajala, Divya
  Tyamagundlu, and Oriana Riva.
\newblock On the effects of data scale on computer control agents.
\newblock {\em arXiv preprint arXiv:2406.03679}, 2024.

\bibitem{christianos2025lightweight}
Filippos Christianos, Georgios Papoudakis, Thomas Coste, Jianye Hao, Jun Wang,
  and Kun Shao.
\newblock Lightweight neural app control.
\newblock {\em arXiv preprint arXiv:2410.17883}, 2024.

\bibitem{wang2025distrl}
Taiyi Wang, Zhihao Wu, Jianheng Liu, Jianye Hao, Jun Wang, and Kun Shao.
\newblock Distrl: An asynchronous distributed reinforcement learning framework
  for on-device control agents.
\newblock {\em arXiv preprint arXiv:2410.14803}, 2024.

\bibitem{agentq}
Pranav Putta, Edmund Mills, Naman Garg, Sumeet Motwani, Chelsea Finn, Divyansh
  Garg, and Rafael Rafailov.
\newblock Agent q: Advanced reasoning and learning for autonomous ai agents.
\newblock {\em arXiv preprint arXiv:2408.07199}, 2024.

\bibitem{koh2024tree}
Jing~Yu Koh, Stephen McAleer, Daniel Fried, and Ruslan Salakhutdinov.
\newblock Tree search for language model agents.
\newblock {\em arXiv preprint arXiv:2407.01476}, 2024.

\bibitem{lecun2022path}
Yann LeCun.
\newblock A path towards autonomous machine intelligence version 0.9. 2,
  2022-06-27.
\newblock {\em Open Review}, 62(1):1--62, 2022.

\bibitem{ding2024understanding}
Jingtao Ding, Yunke Zhang, Yu~Shang, Yuheng Zhang, Zefang Zong, Jie Feng, Yuan
  Yuan, Hongyuan Su, Nian Li, Nicholas Sukiennik, et~al.
\newblock Understanding world or predicting future? a comprehensive survey of
  world models.
\newblock {\em arXiv preprint arXiv:2411.14499}, 2024.

\bibitem{sora}
OpenAI.
\newblock Sora: Creating video from text., 2024.

\bibitem{gameeng}
Dani Valevski, Yaniv Leviathan, Moab Arar, and Shlomi Fruchter.
\newblock Diffusion models are real-time game engines.
\newblock {\em arXiv preprint arXiv:2408.14837}, 2024.

\bibitem{pascanu2017learning}
Razvan Pascanu, Yujia Li, Oriol Vinyals, Nicolas Heess, Lars Buesing, Sebastien
  Racani{\`e}re, David Reichert, Th{\'e}ophane Weber, Daan Wierstra, and Peter
  Battaglia.
\newblock Learning model-based planning from scratch.
\newblock {\em arXiv preprint arXiv:1707.06170}, 2017.

\bibitem{yang2024evaluating}
Chang Yang, Xinrun Wang, Junzhe Jiang, Qinggang Zhang, and Xiao Huang.
\newblock Evaluating world models with llm for decision making.
\newblock {\em arXiv preprint arXiv:2411.08794}, 2024.

\bibitem{go}
Julian Schrittwieser, Ioannis Antonoglou, Thomas Hubert, Karen Simonyan,
  Laurent Sifre, Simon Schmitt, Arthur Guez, Edward Lockhart, Demis Hassabis,
  Thore Graepel, et~al.
\newblock Mastering atari, go, chess and shogi by planning with a learned
  model.
\newblock {\em Nature}, 588(7839):604--609, 2020.

\bibitem{dreamer}
Danijar Hafner, Timothy Lillicrap, Jimmy Ba, and Mohammad Norouzi.
\newblock Dream to control: Learning behaviors by latent imagination.
\newblock {\em arXiv preprint arXiv:1912.01603}, 2019.

\bibitem{liu2023picture}
Anthony Liu, Lajanugen Logeswaran, Sungryull Sohn, and Honglak Lee.
\newblock A picture is worth a thousand words: Language models plan from
  pixels.
\newblock In {\em Proceedings of the 2023 Conference on Empirical Methods in
  Natural Language Processing}, pages 16450--16459, 2023.

\bibitem{customdiffusion}
Nupur Kumari, Bingliang Zhang, Richard Zhang, Eli Shechtman, and Jun-Yan Zhu.
\newblock Multi-concept customization of text-to-image diffusion.
\newblock In {\em CVPR}, pages 1931--1941, 2023.

\bibitem{caoyu}
Yu~Cao and Shaogang Gong.
\newblock Few-shot image generation by conditional relaxing diffusion
  inversion.
\newblock In {\em ECCV}, pages 20--37. Springer, 2024.

\bibitem{layoutlu2023ui}
Yuwen Lu, Ziang Tong, Qinyi Zhao, Chengzhi Zhang, and Toby Jia-Jun Li.
\newblock Ui layout generation with llms guided by ui grammar.
\newblock {\em arXiv preprint arXiv:2310.15455}, 2023.

\bibitem{layoutzheng2023layoutdiffusion}
Guangcong Zheng, Xianpan Zhou, Xuewei Li, Zhongang Qi, Ying Shan, and Xi~Li.
\newblock Layoutdiffusion: Controllable diffusion model for layout-to-image
  generation.
\newblock In {\em CVPR}, pages 22490--22499, 2023.

\bibitem{layoutsobolevsky2023guilget}
Andrey Sobolevsky, Guillaume-Alexandre Bilodeau, Jinghui Cheng, and Jin~LC Guo.
\newblock Guilget: Gui layout generation with transformer.
\newblock {\em arXiv preprint arXiv:2304.09012}, 2023.

\bibitem{layoutzhao2019image}
Bo~Zhao, Lili Meng, Weidong Yin, and Leonid Sigal.
\newblock Image generation from layout.
\newblock In {\em CVPR}, pages 8584--8593, 2019.

\bibitem{text-diffuser}
Jingye Chen, Yupan Huang, Tengchao Lv, Lei Cui, Qifeng Chen, and Furu Wei.
\newblock Textdiffuser: Diffusion models as text painters.
\newblock {\em NeurIPS}, 36, 2024.

\bibitem{zeng2024textctrl}
Weichao Zeng, Yan Shu, Zhenhang Li, Dongbao Yang, and Yu~Zhou.
\newblock Textctrl: Diffusion-based scene text editing with prior guidance
  control.
\newblock {\em arXiv preprint arXiv:2410.10133}, 2024.

\bibitem{paddleocr}
Baoguang Shi, Xiang Bai, and Cong Yao.
\newblock An end-to-end trainable neural network for image-based sequence
  recognition and its application to scene text recognition.
\newblock {\em IEEE transactions on pattern analysis and machine intelligence},
  39(11):2298--2304, 2016.

\bibitem{qiao2020seed}
Zhi Qiao, Yu~Zhou, Dongbao Yang, Yucan Zhou, and Weiping Wang.
\newblock Seed: Semantics enhanced encoder-decoder framework for scene text
  recognition.
\newblock In {\em CVPR}, pages 13528--13537, 2020.

\bibitem{vae}
Diederik~P Kingma and Max Welling.
\newblock Auto-encoding variational bayes.
\newblock {\em arXiv preprint arXiv:1312.6114}, 2013.

\bibitem{unet}
Olaf Ronneberger, Philipp Fischer, and Thomas Brox.
\newblock U-net: Convolutional networks for biomedical image segmentation.
\newblock In {\em Medical image computing and computer-assisted
  intervention--MICCAI 2015: 18th international conference, Munich, Germany,
  October 5-9, 2015, proceedings, part III 18}, pages 234--241. Springer, 2015.

\bibitem{gpt4o}
Aaron Hurst, Adam Lerer, Adam~P Goucher, Adam Perelman, Aditya Ramesh, Aidan
  Clark, AJ~Ostrow, Akila Welihinda, Alan Hayes, Alec Radford, et~al.
\newblock Gpt-4o system card.
\newblock {\em arXiv preprint arXiv:2410.21276}, 2024.

\bibitem{DINO}
Mathilde Caron, Hugo Touvron, Ishan Misra, Herv{\'e} J{\'e}gou, Julien Mairal,
  Piotr Bojanowski, and Armand Joulin.
\newblock Emerging properties in self-supervised vision transformers.
\newblock In {\em CVPR}, pages 9650--9660, 2021.

\bibitem{zhang2023appagent}
Chi Zhang, Zhao Yang, Jiaxuan Liu, Yucheng Han, Xin Chen, Zebiao Huang, Bin Fu,
  and Gang Yu.
\newblock Appagent: Multimodal agents as smartphone users.
\newblock {\em arXiv preprint arXiv:2312.13771}, 2023.

\bibitem{gemini2024}
Demis Hassabis and Koray Kavukcuoglu.
\newblock Introducing gemini 2.0: our new ai model for the agentic era,
  December 2024.
\newblock Accessed: 2025-05-04.

\bibitem{deepfake}
Yisroel Mirsky and Wenke Lee.
\newblock The creation and detection of deepfakes: A survey.
\newblock {\em ACM computing surveys (CSUR)}, 54(1):1--41, 2021.

\end{thebibliography}
}

\newpage
\appendix

In this Appendix, we first provide detailed explanations, including prompts related to our methods, descriptions of our STR dataset, and evaluation details. Then, we present additional experimental results.
Finally, we present additional visualisations of our proposed \ourmethod for GUI generation. 

\section{Experimental Details}
 \label{sec:append_detaisl}
\subsection{GUI-text Predictor}
This subsection elaborates on the design and functionality of the GUI-text predictor, summarising its key components and providing a detailed explanation of its underlying processes.

\dz{Given a STR prediction,} the GUI-text predictor \dz{starts} by locating the text symbols.
To be specific, we first detect black borders by identifying black pixels in the BGR colour space, generating a binary mask that indicates whether a pixel is black or not. A pixel is classified as black if its BGR values fall within the range 
[0,0,0] to 
[50,50,50]. Next, we identify rectangular regions within this mask by computing the ratio of the actual contour area to its corresponding bounding rectangle area. If this ratio exceeds 0.8, the region is considered a valid rectangle, allowing us to extract rectangles with black borders.
For these detected regions, we further analyse their internal colour distribution to determine whether they contain the desired white colour. Specifically, we define white pixels as those with BGR values within the range 
[200,200,200] to 
[255,255,255]. If more than 50\% of the pixels within a region fall within this range, the region is classified as a text symbol. Thus, the locations of text symbols are extracted, and we assign a unique identifier (ID) to each symbol through enumeration.

Building on this, we take as inputs the current GUI image $x_k$, an action $a$ to be applied to this image, the predicted STR ( $\text{STR}_{x_{k+1}^a}$), the location and unique ID token of the text symbols in the STR $\mathcal{T}$.   Then we leverage an LLM to predict the text content for each text symbol. The process begins with preprocessing the STR by overlaying the ID token for each text symbol to the corresponding position in the STR image, resulting in a modified representation denoted as $\text{STR}^{ID}_{k+1}$.
Next, we prompt an LLM to identify which text symbols will remain unchanged after the action $a$ (see the prompt in Subsection \ref{appsubsec:p_1}). These symbols are determined to not be affected by the action and have content identical to the previous GUI $x_k$. Based on the resulting ID list, we retrieve the corresponding pixels from the previous GUI $x_k$ based on their location and update the STR representation. The updated image is still referred to as $\text{STR}^{ID}_{k+1}$ for simplicity. 

Subsequently, the LLM is prompted to determine the semantic role of each text symbol by analysing its context (see the prompt in Subsection \ref{appsubsec:p_2}). This semantic information, combined with $\text{STR}^{ID}_{k+1}$, is then used to predict the exact text content of each symbol (see the prompt in Subsection \ref{appsubsec:p_3}).

Finally, to overlay a symbol with its actual text content, we perform the following steps: 1) For a given text symbol's location and corresponding text, the average background colour is computed by the average colour of the area on the edge of text symbol's coordinates;
   2) The text colour is set to either white or black to ensure optimal contrast with the background colour, for better visibility;
   3)
The font size is calculated as the maximum size that allows the text to fit entirely within the boundaries of the text symbol, ensuring optimal use of space and readability.

\subsection{Action Selection}

In practice, our selection model, described in Section~\ref{subsec:mvwm_agent}, identifies the best action in two steps. First, we query an LLM to evaluate all the action options, providing a judgment—either \textit{valid} or \textit{invalid}—and a confidence score for each action (see the prompt in Subsection \ref{appsubsec:p_4}). These judgments are transformed into scores: if an action is judged \textit{valid}, its score equals the confidence; if judged \textit{invalid}, its score is the confidence multiplied by $-1$. This scoring reflects that higher confidence in a \textit{valid} action yields a higher score, while higher confidence in an \textit{invalid} action results in a lower (negative) score. Second, we query the LLM again to select the best action from the two highest-scoring actions (see the prompt in Subsection \ref{appsubsec:p_5}). This step is motivated by our observation that, in over $70\%$ of tasks, the difference between the top two scores is equal to or less than $0.1$, indicating that both are likely optimal. By allowing the LLM to choose between them, we refine the selection beyond simply picking the action with the highest score.

\begin{table}[t]

\centering
\caption{Summarisation of our STR dataset.}
  \setlength{\tabcolsep}{0.9cm}

\begin{tabular}{lcclc}
\toprule
 Split& App&Episode &Image&Instrucion \\ 
 \hline
  Train &19&2853&19010&14852\\
  Val &19&349&2290&1774\\
  Test &19&348&2320&1824\\
  \hline
  All &19&3550&23620&18450\\
  \hline
\end{tabular}
\label{tab:data-sum-appen}
\end{table}

\subsection{Data Collection}
\label{subsec:data}
To ensure the quality and diversity of data samples for each App, while minimising noise, we collected App information from both Android Control \cite{android_control} and Android in the Wild dataset (AITW)~\cite{rawles2024androidinthewild} datasets.
To be specific, out of 15,274 episodes in the Android Control, only 5,697 episodes include the "open\_app" action. From these episodes, we extracted their "app\_name", identifying 758 unique applications. However, only 13 of these Apps had more than 50 samples. To enrich the dataset, we manually collected additional samples for these 13 Apps from the rest of the dataset.
For AITW, we extracted App names by using the package name listed under the "current activity" field. After filtering out the noisy, 11 valid Apps remained.
By combining the overlapping applications from both datasets, we obtained a total of 19 unique Apps. We split our dataset into "Train", "Validation" and "Test" splits, and we summarise our dataset under each split in Table \ref{tab:data-sum-appen}.

Furthermore, we converted action commands into action instructions for AITW with specific prompts in Subsection \ref{appsubsec:p_6}.
We use Paddleocr \cite{paddleocr} for STR generation.

\begin{table}[t]
\centering
\caption{Decision optimisation comparisons on APP agent performance. Apps are categorised into "Leisure", "Work", and "System". 
}
\setlength{\tabcolsep}{4mm}

\resizebox{1\columnwidth}{!}{
\begin{tabular}{l|l|lccc|c}
        \toprule
App Agent& World Model Modality&World Model &Leisure& Work &System & Overall\\
\hline
\multirow{7}{*}{T3A}
&w/o world model&w/o world model&41.22&51.28& 42.86&43.13\\
\cline{2-7}
&\multirow{2}{*}{Langugae} &Change-text&49.32&51.28& 42.06&46.65 \\
& &HTML-text&47.30&48.72& 43.65&46.01\\
\cline{2-7}
&\multirow{4}{*}{Vision}&HTML-vision&50.68&53.85& 43.65&48.24\\
&&UI-diffuser&48.65&53.85& 43.65&47.28\\
&&IP2P&48.65&53.85& \textbf{45.24}&47.92\\
\cline{3-7}
&&\textbf{\ourmethod} (\textbf{Ours})&50.00&58.97&\textbf{45.24}&49.20\\
\hline
\multirow{7}{*}{APP-Agnet}
&w/o world model&w/o world model&43.24&51.28&39.68&42.81\\
\cline{2-7}
&\multirow{2}{*}{Langugae} &Change-text& 45.96&56.41 &\textbf{45.24}  &46.96  \\
& &HTML-text& 44.59& 56.41& \textbf{45.24} & 46.33\\
\cline{2-7}
&\multirow{4}{*}{Vision}&HTML-vision& 47.97& 56.41& 46.03& 48.24\\
&&UI-diffuser& 47.30& 56.41& 44.44&47.28\\
&&IP2P&47.30 & 58.97& \textbf{45.24}&47.92\\
\cline{3-7}
&&\textbf{\ourmethod} (\textbf{Ours})&50.68&58.97&43.65&48.89\\
\hline
\multirow{7}{*}{Mobile-Agent-v2}
&w/o world model&w/o world model&43.92&53.85&39.68&43.45\\
\cline{2-7}
&\multirow{2}{*}{Langugae} &Change-text&47.30& \textbf{66.67}& 41.27&47.28\\
& &HTML-text&47.30& \textbf{66.67}& 38.89&46.33\\
\cline{2-7}
&\multirow{4}{*}{Vision}&HTML-vision& 50.00& \textbf{66.67}& 41.27& 48.56\\
&&UI-diffuser&49.32 & 61.54&  41.27& 47.60\\
&&IP2P&46.62 &\textbf{66.67} & \textbf{45.24}& 48.56\\
\cline{3-7}
&&\textbf{\ourmethod} (\textbf{Ours})&50.00&\textbf{66.67}&44.44&49.84\\
\hline
\multirow{7}{*}{M3A}
&w/o world model&w/o world model&46.62&51.28& 43.65&46.01\\
\cline{2-7}
&\multirow{2}{*}{Langugae} &Change-text&51.35&51.28& 41.27&47.28\\
& &HTML-text&50.68& 51.28&40.48&46.65\\
\cline{2-7}
&\multirow{4}{*}{Vision}&HTML-vision&52.03&48.72&\textbf{45.24}&48.89 \\
&&UI-diffuser&50.00&48.72&44.44&47.60\\
&&IP2P&52.03&48.72&44.44&48.56\\
\cline{3-7}
&&\textbf{\ourmethod} (\textbf{Ours})&\textbf{53.38}&53.85& \textbf{45.24}&\textbf{50.16}\\
\bottomrule
\end{tabular}
}
\label{table_appendix:breakdown_comre_app_mv2}
\end{table}

\subsection{Evaluation}
\label{subsec: eva}

\paragraph{\dz{World Model Ability.}} For the results \dz{under automatic metrics presented }in Table  \ref{tab:quanti_com}, we prompt LLM for the instructional accuracy score $s_{ia}$ and action readiness score $s_{ar}$, as shown in Subsection~\ref{appsubsec:p_ia} and Subsection \ref{appsubsec:p_ar}  respectively. A generation is considered successful if "success" appears under "Status" for $s_{ia}$ and "yes" under "ready for action"  for $s_{ar}$. \dz{For the user study, we collected 80 generated samples—20 from each of the four world models. We then asked 70 participants to answer three questions on each sample designed to reflect the $s_{ia}$, $s_{gc}$ and $s_{ar}$ scores, as detailed in Subsection  \ref{appsubsec:user_ins}. For the $s_{gc}$, participants are asked to rate on a scale from 1 to 5. These scores were then normalised to  the 
[0,1] range in Table~\ref{tab:quanti_com}.}

\paragraph{\dz{World Model Enhanced App Agent.}}
In Table~\ref{table:agent_com}, we categorised APPs based on their primary functions into three groups: \textbf{Leisure}, \textbf{Work}, and \textbf{System}. The \textbf{Leisure} category includes APPs commonly used for relaxation and entertainment, such as \textit{Decathlon}, \textit{eBay}, \textit{Flipkart}, \textit{Amazon}, \textit{Adidas}, \textit{Kitchen Stories}, \textit{Booking.com}, \textit{YouTube}, and \textit{Vimeo}. The \textbf{Work} category comprises APPs typically associated with professional or productivity-related activities, including \textit{Gmail}, \textit{Drive}, and \textit{Chrome}. Lastly, the \textit{System} category encompasses APPs pre-installed in the Android operating system, such as \textit{com.android.contacts}, \textit{com.google.android.dialer}, \textit{com.google.android.googlequicksearchbox}, \textit{com.android.settings}, \textit{com.google.android.APPs.maps}, and \textit{com.android.vending.}

\paragraph{\dz{Ablation on Iteration Numbers.}} \ourmethod predicts future GUI observations, which can be recursively fed back as input to simulate
further into the future. Taking the generative GUI as the current GUI, an agent was prompted to generate the
action instructions based on the user goal (see the prompt in Subsection \ref{appsubsec:p_7}). Then the action instruction and the GUI were fed into \ourmethod to generate the next GUI. In this study, we defined the iteration number as the number of times \ourmethod was called.  We only use the final output as the signals during the candidate action selection phase, guiding the final selection among potential actions.

\section{Additional Experimental Results}
\dz{\paragraph{Comparison with World Models.}
        Table~\ref{table:compare_wm} compares our \ourmethod with existing world models under M3A App agent. To further highlight our superiority, Table~\ref{table_appendix:breakdown_comre_app_mv2} presents additional results of \ourmethod applied to T3A, APP-Agent, and Mobile-Agent-V2. One can see from the results that our method consistently outperforms existing world models, clearly demonstrating its superiority.}
\begin{table}[t]
\centering
\setlength{\tabcolsep}{8mm}

\caption{Trajectory synthesis evaluation. "T+L" denotes the accuracy of the whole trajectory with length L. }
\begin{tabular}{lcccccc}
        \toprule

World Model &T+1 &T+2 & T+3 &T+4 \\
\hline

w/o world model& 22.81& 14.04&7.02&0\\
\hline
Change-text&52.63&26.32&10.53&5.26\\

HTML-text&38.60&14.04&12.28&7.02\\
HTML-vision&43.86&19.30&10.53&10.53
\\

UI-diffuser&52.63&29.82 & 12.28&5.26\\
 IP2P* &56.14  & 21.05&10.53&7.02\\
\hline
\textbf{\ourmethod} (Ours)&\textbf{57.89}  &\textbf{36.84}& \textbf{14.03}&\textbf{12.28}\\

\bottomrule
\end{tabular}
\label{table: trajectory}

\end{table}
\begin{table}[t]
\centering
\caption{ Evaluation on randomness by running the experiment 3 times (r1-r3) on our sampled test split. "All" denotes the evaluation of the full test split. $s_{gc}$,$s_{ia}$ and $s_{ar}$ are the metrics same with Table \ref{tab:quanti_com}. $s_h$ denotes their harmonic score. STD denote the standard deviation from r1 to r3.
}
\label{tab:random}
\begin{tabular}{c|cccc}
        \toprule
World Model&$s_{gc}$&$s_{ia}$&$s_{ar}$&$s_h$\\
\hline
r1 &0.7421&75.08&78.29&0.7582\\
r2 &0.7323&75.63& 77.64&0.7546\\
r3 &0.7423 &75.39&78.68&0.7605\\
STD & 0.0057 &0.23&0.42&0.0025\\
\hline
ALL& $0.7389$ & $75.37 $ & $78.20 $ & $0.7578 $ \\
\hline

\end{tabular}
\end{table}

\dz{
\textbf{Generation Error Analysis.} As discussed in Subsec. \ref{subsec: ablation}, our method \ourmethod can iteratively generate future GUIs. However, as the number of iterations increases, the accumulated error also grows. In addition to the evidence presented in Table \ref{table:steps}, we conduct further experiments to analyse this iteration error and compare our approach with existing world models. 

To this end, we design a trajectory synthesis evaluation to assess how well the GUIs generated by \ourmethod align with those observations in real-world environments over longer iterations. In this setup, the generated GUI is leveraged as the input to an App agent to generate the subsequent action, with higher-quality trajectories indicating a GUI more aligned with the real-world environment.  Specifically, the GUIs generated by \ourmethod serve as the observation input for the App agent, which generates actions aimed at achieving the user’s goal. These output actions are then evaluated to reflect whether the GUI representations offer concrete and reliable information for action prediction. This process is repeated for 
L steps, and we calculate the success rate of the entire 
L-step trajectory.

 We employ an LLM as a judge to assess the alignment between the agent’s simulated actions and the ground truth actions within a given trajectory. Specifically, an agent was prompted to generate the action instructions based on the given GUI and the user goal (see the prompt in Subsection \ref{appsubsec:p_7}) and the LLM evaluated whether the simulated action lead to the same outcome as the ground truth action (see the prompt in Subsection \ref{appsubsec:p_8}), a "yes" of the "Status" is calculated as a match.

As shown in Table~\ref{table: trajectory},  we compared  \ourmethod against both visual- and language-based world models and demonstrated that while performance decreases across all world models with more iterations,  our model significantly outperformed the other methods by providing more accurate and reliable information. This was reflected in higher trajectory prediction accuracy, underscoring the ability of our model to generate GUIs that aligned with the real-world environment.

}

\dz{
\paragraph{Randomness Study and Evaluation on Full Test Split.}

\ourmethod involves random factors, particularly from the use of LLMs. To evaluate their influence, we conducted the experiment three times, as summarised in Table \ref{tab:random} (r1-r3). The results demonstrate that the randomness does not significantly impact the performance or consistency of our method. Additionally, we focused on a randomly selected subset of examples for evaluation, with results from the full test set also included to illustrate that the observed differences are minor, as shown in Table \ref{tab:random} (compare ALL to r1-r3). We consider the subset results to provide an accurate and reliable approximation for our analysis.
}

\begin{figure}[t]
    \centering
    \includegraphics[width=1\textwidth]{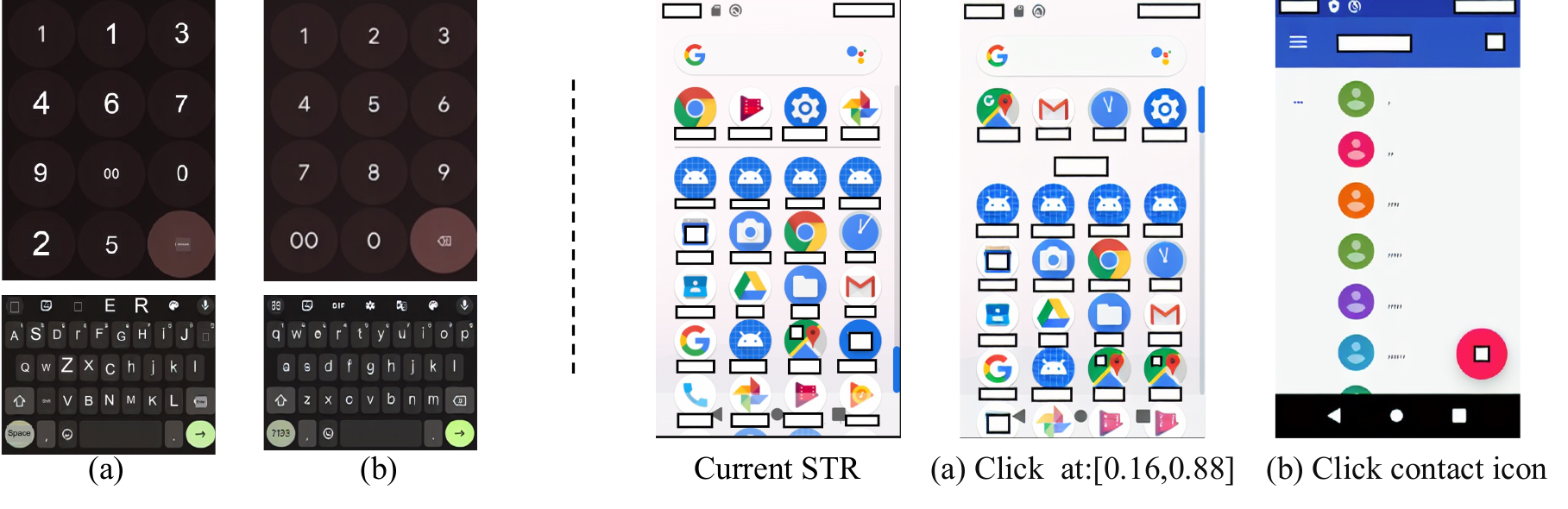} 
    \caption{Qualitative ablation studies. Left: Static text generation. (a) Generating static text via an LLM; (b) Preserving the original text in the image by rendering it as image pixels. Right: STR generation under two input formats—(a) action command and (b) action instruction.}
    \label{fig:appendx_ab}
\end{figure}

\begin{figure}[t]
    \centering
    \includegraphics[width=0.6\textwidth]{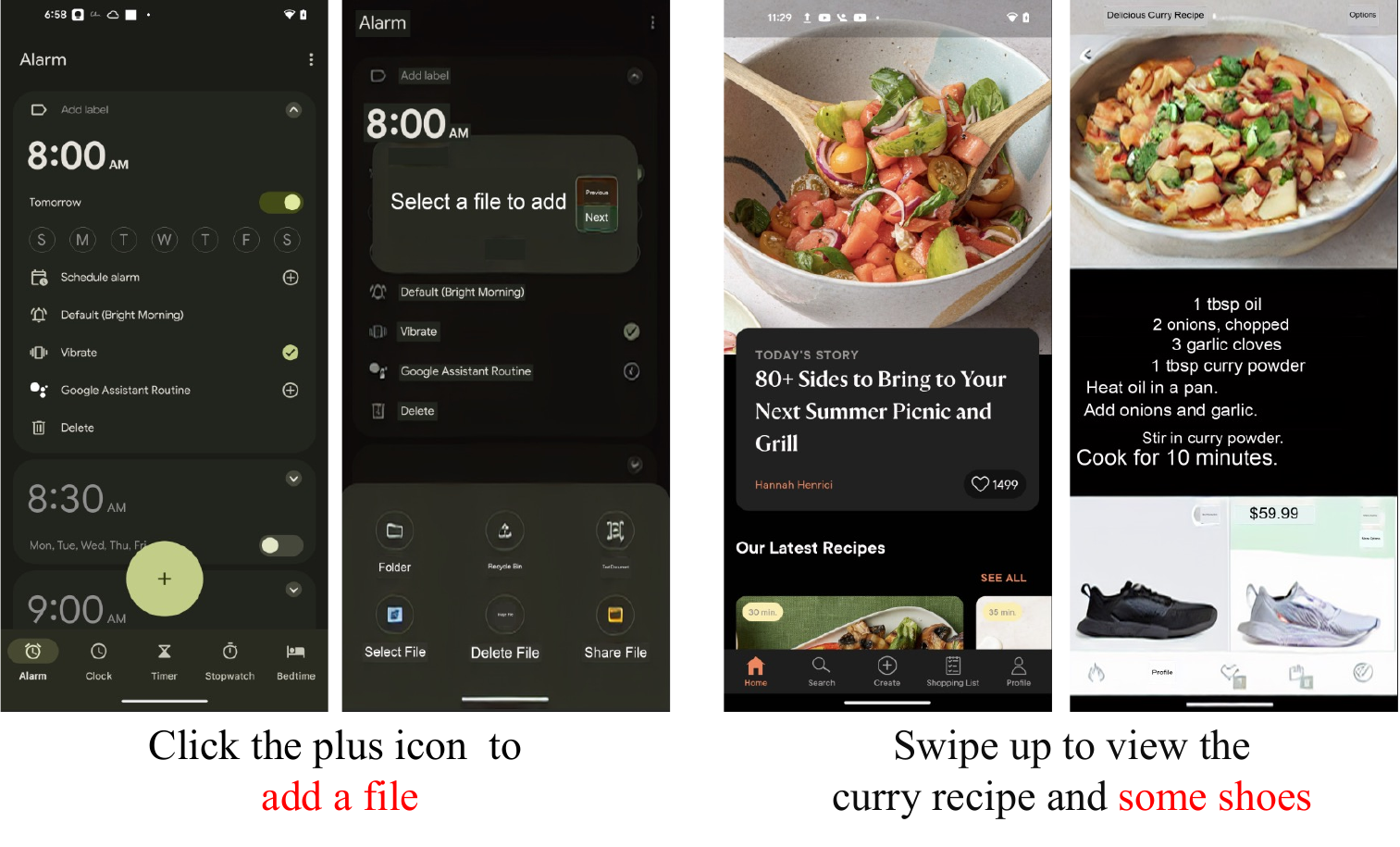} 
    \caption{GUI generation conditioned on a novel combination of current GUI observation and user action.}
    \label{fig:ood_gen}
\end{figure}
\dz{

\paragraph{Qualitative Ablation Analysis.} In addition to the quantitative ablation results presented in Table~\ref{tab:ablation_two}, we also provide qualitative comparisons. Fig.~\ref{fig:appendx_ab} (left) illustrates the challenges faced by the LLM in predicting static text under complex spatial layouts. Fig.~\ref{fig:appendx_ab} (right)  displays the STR generation of the same user intent but with different action types. 
It demonstrated that models learned with action commands failed to predict STR that aligns with the user's intent, whereas action instructions offered a more concrete description, enabling the model to better capture the intent.
}

\paragraph{Qualitative Generalisation Study.}
\dz{We studied the generalisation of \ourmethod in Fig.~\ref{fig:ood_gen} by providing user actions that were not typically encountered within the App's standard context. For example, in the Clock App, a user action to "add a file" generated a Drive-style file selection window while retaining the Clock interface. Similarly, in the Kitchen Store App, \ourmethod can generate content corresponding to the action.  These results emphasised \ourmethod's generalisation ability facing novel combinations of App observations and user actions.
}

\section{\dz{Practical Deployment}}
\label{sec: practical}
In this section, we report the computational efficiency of our method to demonstrate its practicality in real-world applications. The minimum hardware requirement is a GPU with 16 GB of memory. On a V100 GPU, STR image generation takes approximately 8 seconds, and GUI-text prediction takes around 30 seconds. In our setup, the total inference time per request is about 2 minutes, including model loading and communication overhead. Each request involves predicting future GUIs for three different user actions. Table~\ref{tab:inference_time} compares the inference time and step accuracy of \ourmethod with other world models. With an additional 2 minutes of inference time, \ourmethod achieves a notable accuracy improvement of 6.07\%.

\begin{table}[t]
\centering
\caption{Inference time and step accuracy comparison across models.}
\begin{tabular}{lcc}
\toprule
\textbf{Model} & \textbf{Inference Time} & \textbf{Step Accuracy (\%)} \\
\midrule
Baseline (T3A) & $\sim$4 minutes & 43.13 \\
\hline
Change-text    & $\sim$5 seconds & 46.64 \\
IP2P*           & $\sim$1.5 minutes & 47.92 \\
\ourmethod (Ours)    & $\sim$2 minutes & \textbf{49.20} \\
\bottomrule
\end{tabular}
\label{tab:inference_time}
\end{table}
\begin{figure*}[t]
    \centering
\includegraphics[width=1\textwidth]{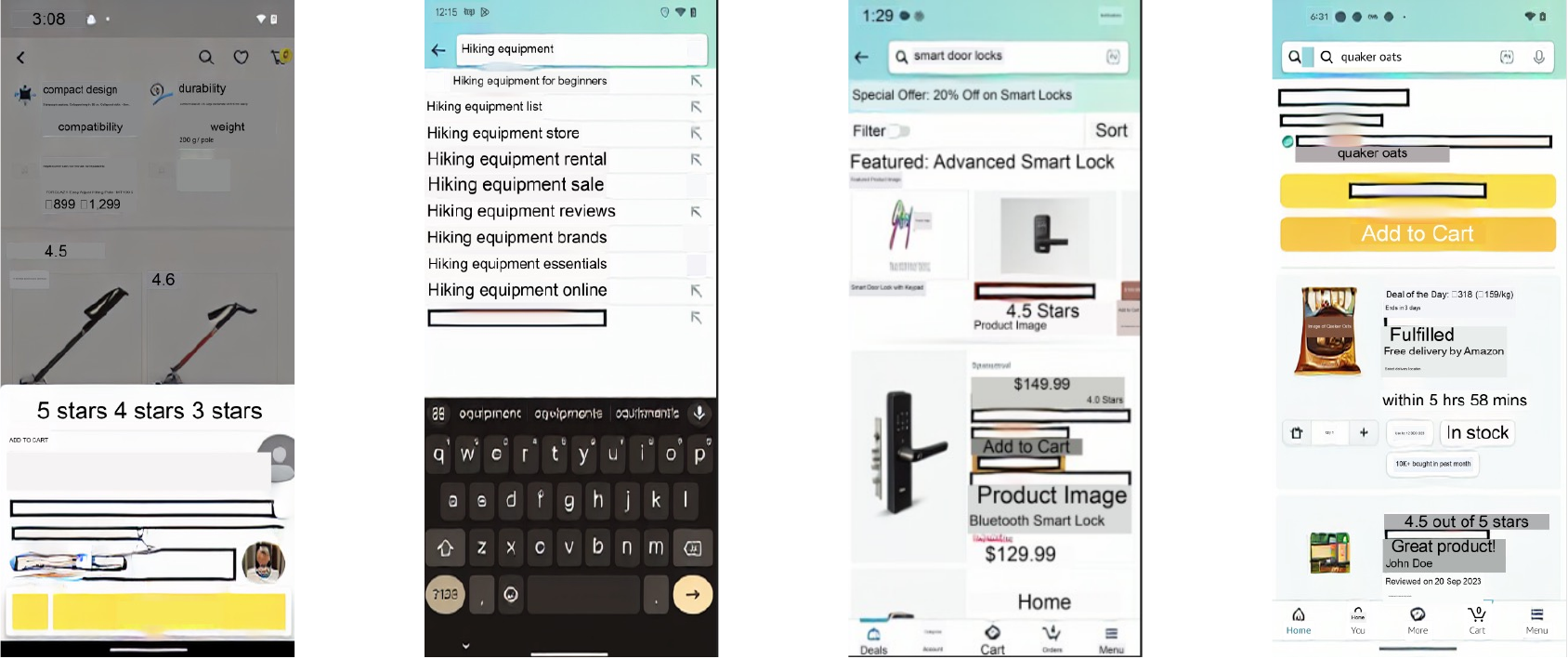}

    \caption{False examples where the text symbols are incorrectly represented, making them unrecognizable to indicate the location of text.}
    \label{fig:appendix_vis_4}
\end{figure*}

\begin{figure*}[t]
    \centering
\includegraphics[width=1\textwidth]{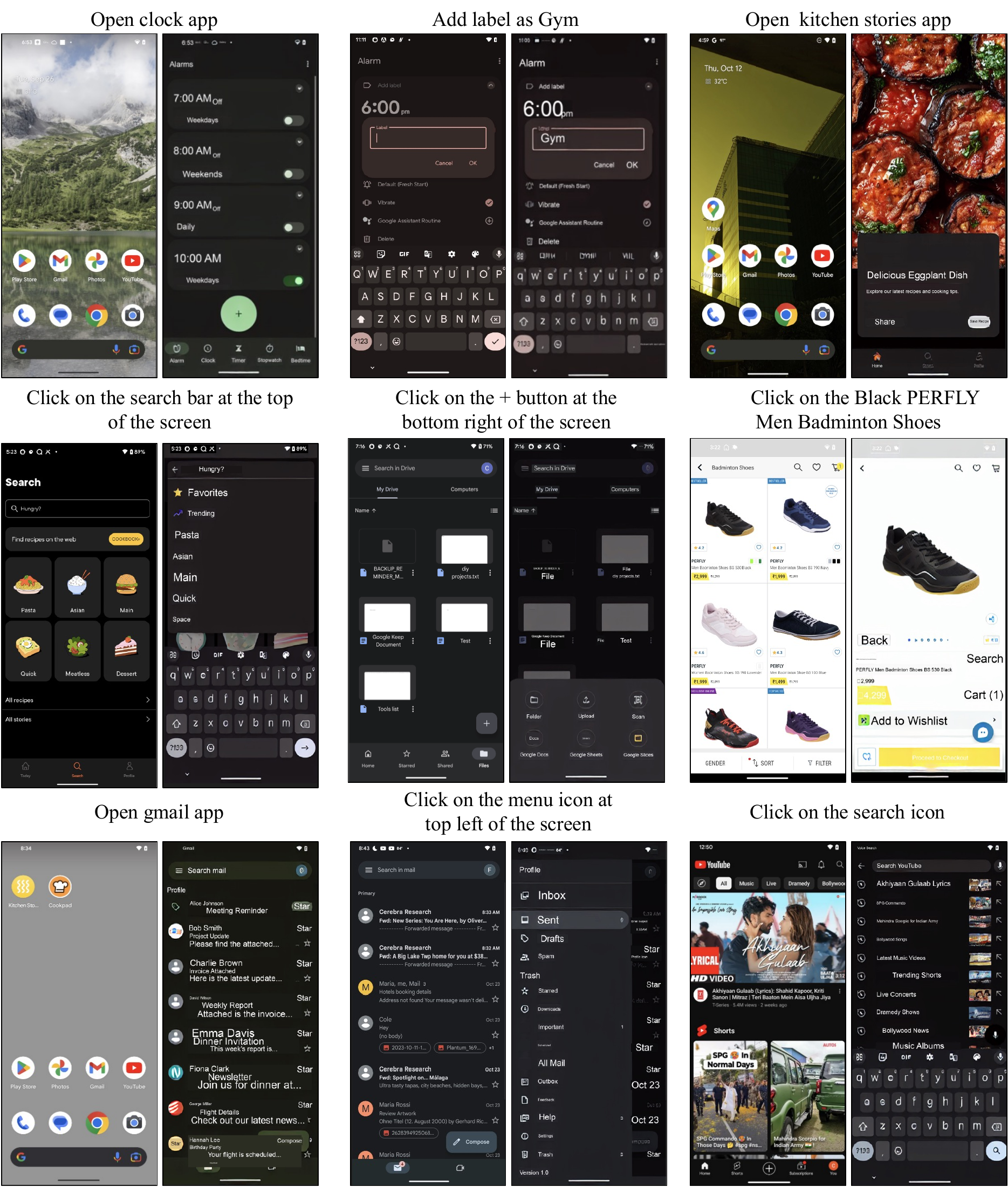}

    \caption{Visualisation of \ourmethod in generating the next GUI. For each example, the action is displayed at the top, with the current GUI shown on the left and the generated GUI on the right.}
    \label{fig:appendix_vis_1}
\end{figure*}

\begin{figure*}[!ht]
    \centering
\includegraphics[width=\textwidth]{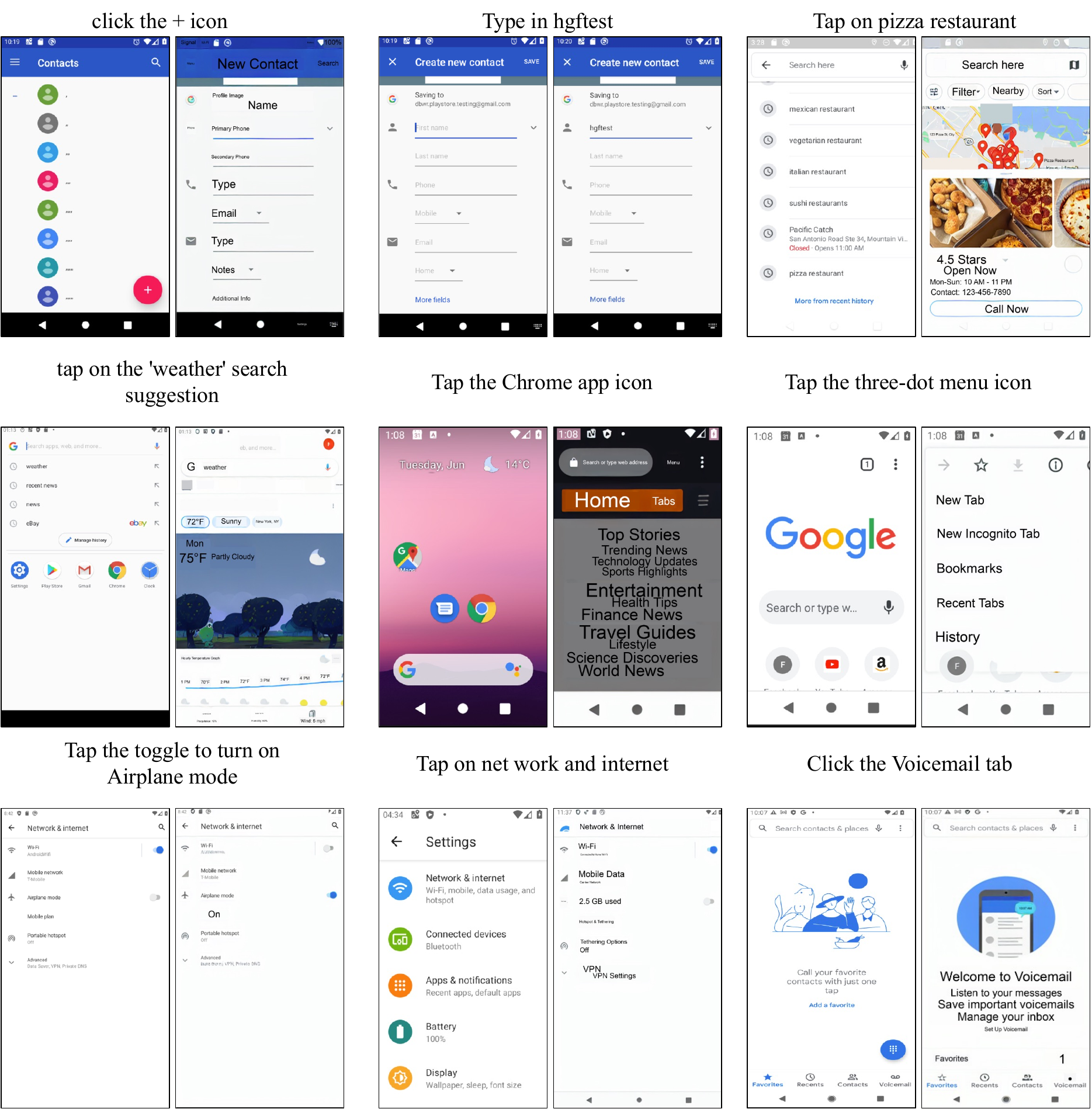}

    \caption{Visualisation of \ourmethod in generating the next GUI. For each example, the action is displayed at the top, with the current GUI shown on the left and the generated GUI on the right.}
    \label{fig:appendix_vis_2}
\end{figure*}
\begin{figure*}[t]
    \centering
\includegraphics[width=1\textwidth]{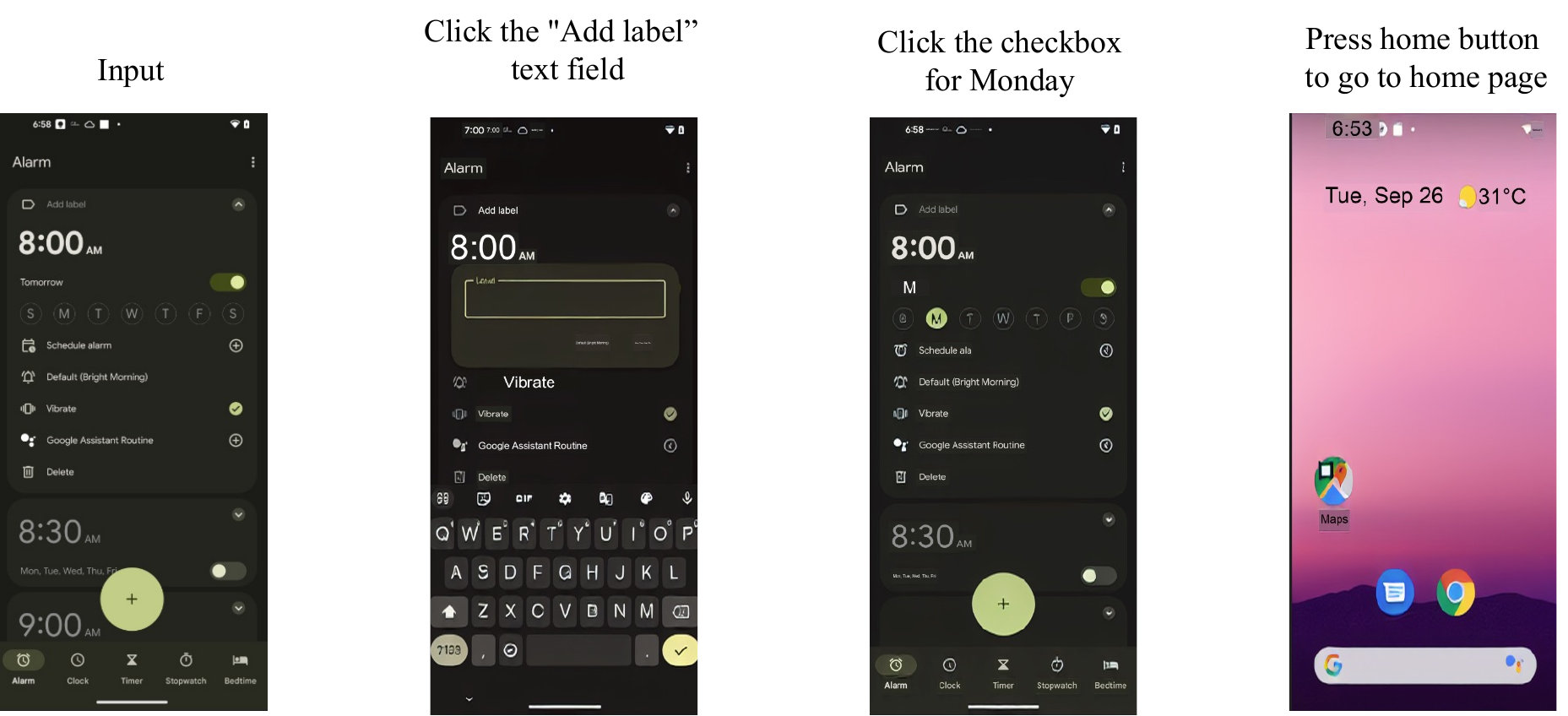}
    \caption{Visualisation of \ourmethod in generating GUIs given a single
current GUI paired with different actions.}
    \label{fig:appendix_vis_3}
\end{figure*}

\section{Limitation}
\label{sec:limi}
Fig. \ref{fig:appendix_vis_4} illustrates failure cases where text symbols are not represented as our rectangle-shaped placeholders with a black border and white fill, making them unrecognisable as text symbols. Improving the representation of text symbols remains a potential direction for future work.

\section{Additional Visualisation}

Diverse visualisations are presented in Fig. \ref{fig:appendix_vis_1} and Fig. \ref{fig:appendix_vis_2}. These examples illustrate how our method effectively generates the next GUI based on the given action and current GUI observation, showcasing its ability to produce visually coherent and contextually accurate GUI simulations.
Moreover, Fig.~\ref{fig:appendix_vis_3} showcases results generated from a single current GUI paired with different actions, further highlighting the versatility of our approach.

\section{Social Impact and Safeguards}
\label{sec: social_impa}
Our work, which focuses on predicting future GUI states, is aligned with general advances in the field of machine learning. As such, it carries potential societal implications similar to those associated with generative technologies, such as deepfakes \cite{deepfake}, including concerns around misuse, misinformation, or user manipulation. To prevent the high risk of misuse of the proposed method, an additional user commitment will be required for accessing the checkpoint in our forthcoming open release, through which we hope to alleviate the potential misuse while benefiting further research.

\section{Prompts}
\subsection{Prompt to decide the text symbols to remain unchanged after the action}
\label{appsubsec:p_1}
\begin{tcolorbox}[mypromptstyle]
You are a professional UI/UX analyst and your goal is to compare the two UI screenshots and return their overlapping layout.

\textit{\#\#\#} Inputs:

1. **Current Screenshot**: The current mobile UI as an image.

2. **Next UI Layout Screenshot**: 

- An image of the next mobile UI layout with all text replaced by white boxes.

- Each box has a unique red ID label.

3. Use action: a user action described by language

Next UI Layout Screenshot is a result of a user action on the current screenshot, but the text elements are masked.

Please help me identify those layouts that are located in the same position, so I can predict their text directly from the current screenshot.

Usually, the system bar information should be included.
Exclude elements from the result if:

The content (text) changes as a result of the user action, even if the element exists in both screenshots.

Please be very very cautious about putting an ID on the list, which means you are very very confident with this task. if you are unsure about some elements, please ignore them and do not put them on the list.

\textit{\#\#\#} Output the list of existing elements :
Return the results in the following JSON format:
['id1','id2',...]

\textit{\#\#\#}  Notes:

- Ensure the detected elements appear in both UI screenshots, which means their surrounding context is the same.

- Ensure identify those elements that their text will change by the user action and exclude them from your response.

- Ensure identify those elements that share a similar context layout, but their absolute are not the same, and them from your response.

- Ensure only reply in pure JSON format, with no placeholders or comments.

\end{tcolorbox}

\subsection{Prompt to determine the semantic role of each text symbol}
\label{appsubsec:p_2}
\begin{tcolorbox}[mypromptstyle]

You are a professional UI/UX analyst assigned to structure and analyse the semantics of mobile UI screenshots. 

Your goal is to segment the UI and annotate box elements in a way that enhances understanding of their roles and relationships within the interface. 
Inputs: 

- Current Screenshot: A visual representation of the mobile UI. 

- Next UI layout screenshot: A visual representation of the next UI layout with all the text masked with a white box. Each box has an ID number on it in red colour.
-
User Action: An action put on the current UI will result in the next UI. 

- Box locations: a list of box locations to better help you to locate the boxes in the format of {'id': id, 'Location':[x1,y1, width, height]}. ID indicates their ID number in the UI screenshot. 

- UI\_size: the width/height of the input images. They are the same size. The image you received might be resized. Please scale it back for the locations. 

Task: 

Structure the boxes in the Next UI layout screenshot with semantics based on the visual input by following these steps: 

1, Divide the UI into Semantic Windows Group the UI into functional sections with a specific name (e.g., "Header Windows," "Time Selector Panel"). 

2. Structure Text Elements in Each Semantic Window.

- Assign box elements to windows based on logical, visual relationships or semantic roles.

- For every element, structure output as : 

**id: corresponding box retrieved from the box list and the Next UI layout screenshot.

**Role: A brief explanation of the role of this box. You should consider their [x1,y1] to indicate their location, [w,h] to indicate their size to decide the role. It is important to consider the context for the role prediction. The role should be in detail to distinguish it from other items in the same category. 

Output Format: 
\{ "Window Name": { "Category Name": [{ "id":id, "Role": "Role" }, { "id":id, "Role": "Role" }, ... ], "Category Name": [{ "id":id, "Role": "Role"  },  ... ]  },  ... \} 

Key Guidelines:

- Ensure to retrieve id from the given screenshot and box list. 

- Avoid duplicating or omitting IDs.

- Every box element in the box location list must be included in the structured output.

- Ensure there is no additional formatting, code blocks or placeholders in your response; return only a clean JSON without any comments.
\end{tcolorbox}

\subsection{Prompt to predict the exact text content for each symbol}
\label{appsubsec:p_3}
\begin{tcolorbox}[mypromptstyle]
Task: Plan the content for the next UI screen based on the provided inputs and instructions.

Inputs:

Current Screenshot: A visual representation of the mobile UI.

Next UI layout screenshot: A visual representation of the next UI layout with all light yellow boxes indicating a text place. Each box has an ID number on it.

User Instruction: A specific action or command that transitions the current UI to the next UI state.

Semantics for the masks in Next UI screenshot: A structured map.

Goal:

Predict the content (text) for each masked area in the next UI layout screenshot based on the following steps:

Map Affected Elements to the Next UI.

Align the affected elements with the yellow box coordinates on the next UI.

Predict the text for each yellow box based on the user instruction and the context of 
the current UI.

If you can not find any information about the text, predict a plausible text based on its context.

Ensure to use the semantics to help you understand the layout and predict the text. If you think the semantics is wrong, please modify it in your

Output:

Return the predictions in JSON format with the structure:
\{"Window Name ": {"Category Name ": [ {"id ": id, "text ": “text”, "role ": "role" }, {"id ": "id", "text": "text", "role": "role " } ], }, ... \}

Ensure to predict text based on the context.

Do not include any special characters.

Ensure there is no additional formatting, code blocks or placeholders in your response; return only a clean JSON without any comments.
\end{tcolorbox}

\subsection{Prompt to evaluate actions with a confidence score}
\label{appsubsec:p_4}
\begin{tcolorbox}[mypromptstyle]
You are an agent who can operate an Android phone on behalf of a user. When given a user request, you will try to complete it step by step. At each step, a list of descriptions for most UI elements on the current screen will be given to you (each element can be
specified by an index), together with a history of what you have done in previous steps. Based on these pieces of information and
the goal, you must choose to perform one of the actions in the following list (action description followed by the JSON format) by
outputting the action in the correct JSON format:
{action options from the dataset}

The overall user goal/request is: \{goal\}

Here is a history of what you have done so far:\{history\}
This is the action you picked in the latest step: \{action\}, whose semantic description is: \{sum\}

Your goal is to judge **whether the action you picked in the latest step is on the right track to the successful execution of the
overall user goal/request**.

You will be given the screenshots before and after you perform the action

- The first screenshot corresponds to the UI state before you performed the action.

- The second screenshot corresponds to the UI state after you performed the action.

Also here is the list of detailed information for some UI elements in the before screenshot:
\{before\_elements\}

Note that, the "after" screenshot is generated by the agent’s world model.
As such, it may not faithfully represent the real UI. For instance: Some UI elements in the simulated "after" screenshot may not
exist in a real UI.
Your evaluation should consider the reliability of the UI predictions. If the "after" screenshot contains unreasonable elements, this
likely indicates a failure.

Now provide your judgment on the selected action in JSON format. Your response must include:

Reason: A detailed explanation of why the action is valid or invalid.

Judgment: Your judgment must be either "valid" or "invalid".

Confidence: A confidence score between 0.0 and 1.0, reflects how likely your judgment is correct.

You must follow this structure exactly:

\{Reason: ..., Judgement: "valid" or "invalid", Confidence: ...\}

Your Answer:

\end{tcolorbox}

\subsection{Prompt to select the optimal actions among two highest-scoring actions}
\label{appsubsec:p_5}
\begin{tcolorbox}[mypromptstyle]
    You are an agent who can operate an Android phone on behalf of a user. When given a user request, you will try to complete it
step by step. At each step, a list of descriptions for most UI elements on the current screen will be given to you (each element
can be specified by an index), together with a history of what you have done in previous steps. Based on these pieces of
information and the goal, you must choose to perform one of the actions in the following list (action description followed by
the JSON format) by outputting the action in the correct JSON format
{action options from the dataset}

 The overall user goal/request is: \{goal\}

Here is a history of what you have done so far:\{history\}
 
 Here is a list of descriptions for some UI elements on the current screen:\{before\_elements\}

Here are two candidate actions:

Action 1: \{action\_0\}, described semantically as \{sum\_0\}. The rationale for this action is: \{act\_re\_0\}

Action 2: \{action\_1\}, described semantically as \{sum\_1\}. The rationale for this action is: \{act\_re\_1\}

Hints for making your decision:
\{GUIDANCE\}

- Both "more options" buttons and scrolling actions may reveal new content. Evaluate which is more suitable for the goal.

- Consider the history of previous actions. If prior steps involved repeated "scroll down" actions, it is more likely that "scroll
down" is the correct next step.

- If the user goal involves viewing reviews or similar tasks and the current screen already displays such content, "scroll down"
may reveal more information.

Your task is to choose the best action from the two provided.

Now, provide your judgment in JSON format with the following structure:

Reason: A detailed explanation of your choice, considering the hints above.

Choice: Action 1 or Action 2.

Your output must exactly match this format:

\{Reason: ..., Choice: Action 1 or Action 2\}
\end{tcolorbox}

\subsection{Prompt to convert action commands into action instructions}
\label{appsubsec:p_6}
\begin{tcolorbox}[mypromptstyle]
You are a professional UI/UX analyst specializing in identifying the semantics of dual point actions between mobile UI
screenshots.

Inputs:

Current Screenshot: A visual representation of the mobile UI.

Next Screenshot: A visual representation of the NEXT mobile UI.

Goal: A user intent on this Mobile interface.

touch\_xy: the x,y coordinates for the touch point, as a percentage of the image dimensions.

lift\_xy: the x,y coordinates for the lift point, as a percentage of the image dimensions.

Your task is to analyse these elements describe the precise user action in plain language and return your answer in plain
string (e.g., "click the + icon", "scroll up").

If the two screenshots are identical, please return an empty string as "".

If the Next Screenshot does not seem to be one step away from the Current Screenshot, return an empty string as "".
One step means only one interaction with the cell phone.

Ensure there is no additional formatting, code blocks or placeholders in your response; return only a clean string without any
comments
\end{tcolorbox}

\subsection{Prompt for instructional accuracy score~($s_{ia}$)}
\label{appsubsec:p_ia}
\begin{tcolorbox}[mypromptstyle]
    You are an expert in evaluating the performance of a mobile emulator. The mobile emulator is designed to
navigate the UI change based on human instruction.

Inputs:

Current UI Screenshot: The present state of the cellphone's user interface.

Next UI Screenshot: The mobile emulator generated UI indicating the next state of the cellphone's user interface based on human instruction.

Human instruction: The action applied on the current UI screenshot.

Your goal is to determine whether the mobile emulator successfully predicts the next UI image with current information and layout based on the current UI and the user action.  

*IMPORTANT*

Format your response into a JSON map as shown below:

\{

"Thoughts": <your thoughts and 
reasoning process>,

"Status": "success" or "failure",

\}

\end{tcolorbox}

\subsection{Prompt for action readiness accuracy score~($s_{ar}$)}
\label{appsubsec:p_ar}
\begin{tcolorbox}[mypromptstyle]
    You are an expert in evaluating the performance of a mobile emulator. The mobile emulator is designed to
navigate the UI change based on human instruction.

Inputs:

UI Screenshot: The mobile emulator generated UI indicating the state of the cellphone's user interface.

User intent: The user goal to achieve.

Next action: the action will be applied to this UI.

Your goal is to determine whether the next action is validated on the UI Screenshot. 

Please also indicate if it is still in the right App according to the goal.

*IMPORTANT*
Format your response into a JSON map as shown below:

\{

"Thoughts": <your thoughts and reasoning process>,

"In the right App": "yes" or "no"

"ready for action": "yes" or "yes",

\}

\end{tcolorbox}

\begin{figure*}[ht]
    \centering
\includegraphics[width=1\textwidth]{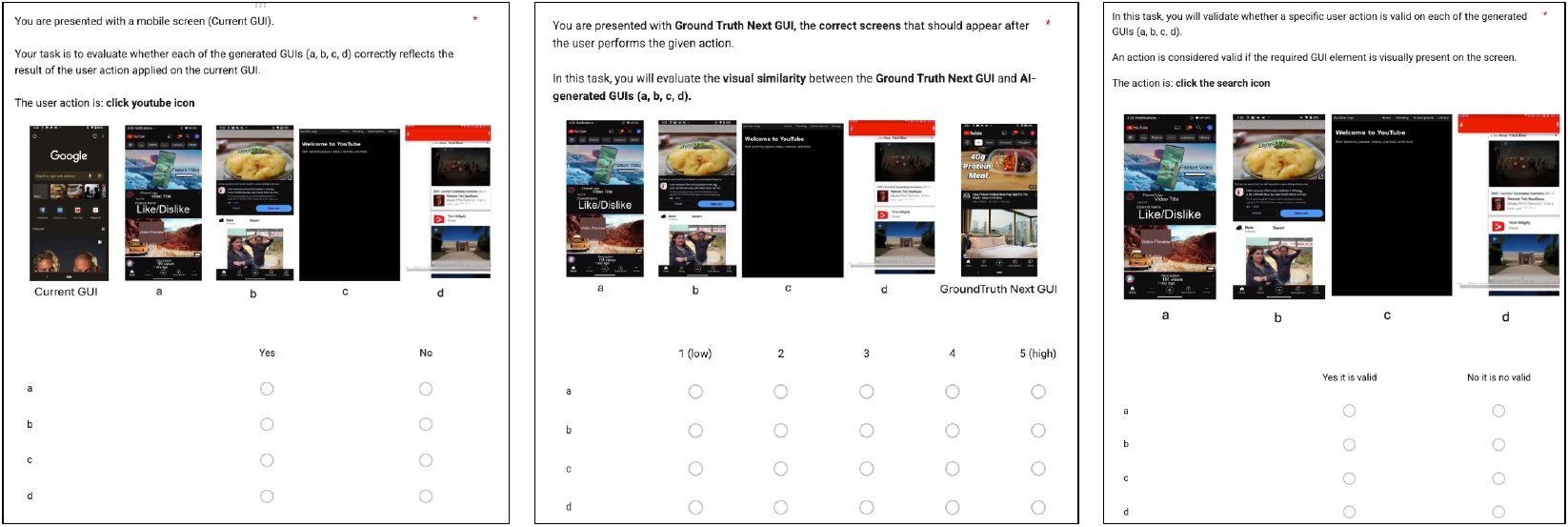}

    \caption{Screenshot of user study example.}
    \label{fig:appendxi_user_study}
\end{figure*}

\subsection{Instructions for User Study}
\label{appsubsec:user_ins}
The following prompt provides the instructions for the user study. An example screenshot is shown in Fig.~\ref{fig:appendxi_user_study}.

\begin{tcolorbox}[mypromptstyle]
Question 1:
You are presented with a mobile screen (Current GUI).

Your task is to evaluate whether the generated GUI correctly reflects the result of the user action applied on the current GUI. Answer "Yes" or "No" to each sample.

Question 2:

You are presented with Ground Truth Next GUI, the correct screens that should appear after the user performs the given action.

In this task, you will evaluate the visual similarity between the Ground Truth Next GUI and AI-generated GUI, scoring from 1-5.

Question 3:

In this task, you will validate whether a specific user action is valid on the generated GUI.

An action is considered valid if the required GUI element is visually presented on the screen. Answer "Yes" or "No" to each sample.

\end{tcolorbox}

\subsection{Prompt to generate the action instruction based on the given GUI and the user goal}
\label{appsubsec:p_7}
\begin{tcolorbox}[mypromptstyle]
You are an autonomous intelligent agent tasked with navigating a cell phone to accomplish specific tasks. You will be provided
with the following information:

1. Initial UI screenshot: A visual representation of the initial state of the cell phone's interface.

2. User Objective: This is the task you are trying to complete.

3. Previous Action: An action sequence performed on the initial UI.

4, Current UI states: A visual representation of the current state of the cell phone's interface, generated by a simulated
environment.

The initial image is the screenshot before actually performing all the previous actions.

The current cell phone UI is generated by applying previous actions on the initial screenshot.

Your Task: Please predict a single next step action to complete the given task based on current vision states.

To be successful, it is very important to follow the following rules:

1. You should only issue one action that is valid based on the current UI states.

2. You should only issue one action at a time. Avoid issuing multiple actions like "do A and do B".

3. Generate the action in plain text. For example, Scroll down to set the minute as 15.

4. Issue "Stop." if you think the action is already completed.
Ensure you only return the action, not other formats, comments or placeholders

\end{tcolorbox}

\subsection{Prompt to evaluates whether the simulated action leads to the same outcome as the ground truth action}
\label{appsubsec:p_8}
\begin{tcolorbox}[mypromptstyle]
You are an expert in evaluating the performance of a cell phone navigation agent. The agent is designed to
help a human user navigate a cellphone to complete a task.

Inputs:

Current UI Screenshot: The present state of the cellphone's user interface.

User Intent: The goal the human user aims to achieve.

Action History: The sequence of actions taken so far for you to track the progress.

Agent Simulated Action: The action suggested by the agent to achieve the user’s intent.

Ground Truth Action: The correct action is needed to achieve the user’s intent.

Your goal is to determine whether the agent’s simulated action leads to the same outcome as the ground truth action.

Additionally,
if the simulated action does not exactly match the ground truth action but is still progressing toward the correct
outcome to achieve user intent, indicating that the action is "on the right track."

*IMPORTANT*

Format your response into a JSON map as shown below:

\{

"Thoughts": <your thoughts and reasoning process>,

"Status": "success" or "failure",

"On the right track to success": "yes" or "no"

\}

\end{tcolorbox}

\end{document}